\documentclass[journal]{IEEEtran}
\usepackage{tikz}

\usepackage{graphicx}
\usepackage{epsfig}
\usepackage{latexsym}
\usepackage{amsfonts}
\usepackage{here}
\usepackage{rawfonts}
\usepackage[latin1]{inputenc}
\usepackage[T1]{fontenc}
\usepackage{calc}
\usepackage{url}
\usepackage{enumerate}
\usepackage{color}
\usepackage[tbtags]{amsmath}
\usepackage{amssymb}
\usepackage{upref}
\usepackage{epic,eepic}
\usepackage{times}
\usepackage{dsfont}
\usepackage{comment}
\usepackage{cite}
\usepackage{algorithm}
\usepackage{algpseudocode}
\usepackage{graphicx}
\usepackage[font={small}]{caption}
\usepackage[font={small}]{subcaption}

\usepackage{stfloats}
\usepackage{mathtools}
\usepackage{amsmath}

\newtheorem{theorem}{\bf{Theorem}}

\newtheorem{corollary}{\bf{Corollary}}

\newtheorem{remark}{Remark}

\ifCLASSINFOpdf
\else
\fi
\begin{document}

\title{Over-the-Air Federated Learning via Weighted Aggregation}

\author{{Seyed Mohammad Azimi-Abarghouyi} and Leandros Tassiulas,~\IEEEmembership{Fellow,~IEEE}
  \thanks{S. M. Azimi-Abarghouyi is with the School of Electrical Engineering and Computer Science, KTH Royal Institute of Technology, Stockholm, Sweden (e-mail: seyaa@kth.se). L. Tassiulas is with the Department of Electrical Engineering, Institute for Network Science, Yale University, New Haven, CT USA (e-mail: leandros.tassiulas@yale.edu).}
}

\maketitle

\begin{abstract}
This paper introduces a new federated learning scheme that leverages over-the-air computation. A novel feature of this scheme is the proposal to employ adaptive weights during aggregation, a facet treated as predefined in other over-the-air schemes. This can mitigate the impact of wireless channel conditions on learning performance, without needing channel state information at transmitter side (CSIT). We provide a mathematical methodology to derive the convergence bound for the proposed scheme in the context of computational heterogeneity and general loss functions, supplemented with design insights. Accordingly, we propose aggregation cost metrics and efficient algorithms to find optimized weights for the aggregation. Finally,
through numerical experiments, we validate the effectiveness of the proposed scheme. Even with the challenges posed by channel conditions and device heterogeneity, the proposed scheme surpasses other over-the-air strategies by an accuracy improvement of $15\%$ over the scheme using CSIT and $30\%$ compared to the one without CSIT.

\end{abstract}

\section{Introduction}
As wireless edge devices such as phones, smart watches, sensors, and autonomous vehicles become more prevalent and powerful, there is a growing need to use machine learning to train a global model--- a unified algorithm that learns patterns from diverse data sources over these devices. However, it is often not possible to transfer such large amounts of data from the devices to a central server due to latency, power, bandwidth, and data privacy concerns. A distributed approach called federated learning (FL) is a promising solution as it enables machine learning to be implemented directly at the wireless edge without transferring data off the devices \cite{mcmahan}. This method can be applied to a variety of applications \cite{vsmith, viktoria}. In FL, the model training is performed locally at each device with the help of a parameter server, and the process of parallel model updates along with aggregation at the server is performed iteratively until it reaches convergence. FL is often conducted over unreliable wireless networks with limited resources and power constraints. In these networks, devices and the edge server communicate via a shared wireless propagation medium, making communication efficiency a critical challenge for FL. Conventional methods, including those using digital communications, segregate communication and computation tasks. These methods typically employ orthogonal multiple access techniques to avoid interference, requiring separate transmissions from each device to the server in distinct resource blocks, leading to significant communication latency and resource requirements \cite{viktoria}. A cost-effective FL scheme is proposed in \cite{leandros} to address these challenges by optimally determining the number of devices to select and the number of local iterations in each training round. FL using non-orthogonal multiple access (NOMA) techniques via digital communications is also proposed in \cite{noma}. While this approach reduces resource usage, the interference acts as additional noise. This is because aggregation is performed only after the individual decoding for each device is complete.

Over-the-air computation \cite{nazer, power_huang} is a promising scheme based on analog communications that leverages the superposition property of wireless channels, allowing simultaneous multiple access transmissions from edge devices within a single resource block. Based on this scheme, an approach known as over-the-air FL is proposed to perform aggregation under the interference, by merging communication and computation. Over-the-air FL can function with remarkably fewer resources and lower latency compared to FL using orthogonal transmissions \cite{alphan, viktoria}. Additionally, over-the-air FL utilizes interference by directly recovering the aggregation function as a whole, without any individual recovery for each device, unlike NOMA approaches \cite{alphan, viktoria}. However, the aggregation process in over-the-air FL is prone to estimation errors due to the severe effects of wireless channel fading on communication links. This research presents a novel approach to tackle these challenges in over-the-air FL.

\subsection{Prior Work}
Most previous research on over-the-air FL has assumed the necessity of perfect channel state information at transmitter side (CSIT) for all the devices. Using a power allocation approach for the purpose of channel compensation, this information aids in adjusting transmission powers and phases, thereby rectifying the misalignment between wireless channels and predefined aggregation weights \cite{gunduz1, gunduz2, cao, aircomp, cao1, azimi_FL, tao, huang_analog, huang_sg, huang_turning, osvaldo, wsaad, go, wind}. However, this approach places a substantial burden on channel estimation training and feedback
mechanisms for each device, leading to increased latency before each transmission and a notable decrease in spectral and energy efficiency. 
It also requires perfect synchronization among the transmitters, each equipped with extra hardware for accurate channel adjustments. Additionally, poor channel conditions may either prevent a device from contributing to the learning process or require it to use substantial transmission power. This situation results in signal distortions and imperfect aggregation due to the inherent physical limitations on the maximum and average power capacities of each device \cite{alphan}. Hence, this approach presents certain challenges when applied to low-cost distributed learning across a large number of devices, particularly those with limited power and hardware capabilities. The common strategy to power allocation is truncated power allocation, where each transmitter only needs to know its own channel, referred to as local CSIT \cite{gunduz1, gunduz2, azimi_FL, huang_analog, huang_sg, huang_turning, osvaldo}. Another strategy involves joint device selection and power allocation schemes, as proposed in \cite{wsaad, aircomp, go, wind}. These studies require global knowledge of all channels at every device before each transmission, referred to as global CSIT, to centralize the allocation optimization process. This differs from the distributed approach of the truncated power allocation. At its core, the device selection strategy aims to include the maximum number of devices in each communication round, as an aggregation metric, ensuring that an estimation cost term remains beneath a tolerable threshold for a predefined aggregation. 
Additionally, there have been power allocation strategies that rely on the presence of knowledge regarding model and data statistics, as referenced in \cite{cao, cao1, tao}. This prerequisite can constrain the utility of over-the-air FL in various applications. 


Many current wireless systems transmit blindly using a constant power. Indeed, apart from eliminating the need for CSIT, there are several advantages to transmitting without explicitly compensating for the channel. Primarily, it facilitates maintaining the average transmission energy of the signal regardless of the channel conditions. Also, it prevents the enlargement of the dynamic range of the transmitted signal without any adaptation for channel compensations, thereby significantly reducing the complexity of hardware implementation requirements. Lastly, efforts to correct the channel at the transmitter could be compromised by channel estimation errors, leading to values at the receiver that are multiplied by unpredictable gains \cite{cohen1}. Taking these into account,
the idea of a blind over-the-air FL approach has become increasingly prominent \cite{alphan}. The studies in \cite{cohen1, cohen2, yang} delve into blind over-the-air FL without any compensation for the detrimental effects of fading. They confirm that despite these factors, convergence is still assured. In \cite{gunduz3, turky, amiri22}, blind over-the-air schemes are presented that leverage multiple antennas and rely on channel state information at the receiver side (CSIR). Nonetheless, these schemes necessitate large multiple receiver antennas, and the effects of wireless fading become less severe as the number of antennas tends toward infinity. In addition, a blind scheme that uses lattice coding at the device, along with a sufficiently large number of receiver antennas and CSIR at the server, is proposed in \cite{azimi_lattice, myisit}. Contrary to the common use of analog modulation in over-the-air FL, this scheme is designed for digital modulation.

\subsection{Key Contributions}
We propose a novel over-the-air FL scheme named weighted over-the-air FL ({\fontfamily{lmtt}\selectfont
	WAFeL}) to counteract the negative impact of wireless channels on the learning convergence performance by leveraging adaptive aggregation weights. Importantly, {\fontfamily{lmtt}\selectfont
	WAFeL} operates as a blind scheme, eliminating the need for CSIT. Without channel compensation at the transmitter end, this approach is shown to be effective in mitigating estimation error of the aggregation caused by different wireless channel conditions of the devices. This remains the case even when deploying a single-antenna server. Our approach uniquely introduces using aggregation weights as optimizable parameters, where channel compensation is one of the objectives addressed. This sets it apart from other blind schemes that either forgo compensation altogether or need an extensive number of antennas for the same. In summary, our proposal offers the following major contributions. 

\textit{Weighted Over-the-Air Scheme:} We propose a generalized approach to aggregation, {\fontfamily{lmtt}\selectfont
	WAFeL}, which differs from the conventional method of using predefined weights, such as equal or proportional to local dataset sizes. The seminal paper on FL \cite{mcmahan} originally presented this aggregation method assuming ideal transmission circumstances and perfect aggregation estimation. However, subsequent research on over-the-air FL has continued to use the same aggregation method, regardless of interference and transmission imperfections \cite{gunduz1, gunduz2, cao, aircomp, cao1, azimi_FL, tao, huang_analog, huang_sg, huang_turning, osvaldo, wsaad, go, wind, gunduz3, cohen1, cohen2, yang,turky,amiri22}. In contrast, our work employs adaptive aggregation weights to mitigate aggregation estimation error and its effects, while a desirable learning performance is ensured. Furthermore, while most previous over-the-air FL research has not incorporated the computational heterogeneity of devices in their scheme design \cite{gunduz1, gunduz2, cao, aircomp, cao1, azimi_FL, tao, huang_analog, huang_sg, huang_turning, osvaldo, wsaad, go, wind, gunduz3, cohen1, cohen2, yang, azimi_lattice, myisit, turky, amiri22}, our approach integrates varying batch sizes across devices based on their computational strengths and a deadline. This is embodied in {\fontfamily{lmtt}\selectfont WAFeL}, resulting in a design that is optimized for device heterogeneity.

\textit{Receiver Architecture:} In order to accomplish the proposed weighted aggregation, we propose a new receiver architecture at the server side that incorporates both the real and imaginary parts of the signal, along with an equalization vector. This is then optimized with the objective of reducing the mean squared error (MSE) of the estimation to a minimum. In other over-the-air schemes \cite{gunduz1, gunduz2, cao, aircomp, cao1, azimi_FL, tao, huang_analog, huang_sg, huang_turning, osvaldo, wsaad, go, wind, cohen1, cohen2, yang}, perfect synchronization ensures that the received signal is real, leading to a different processing and receiver architecture tailored for their equal aggregation weights.

\textit{Convergence Analysis:} Based on a basic set of broadly accepted principles, we analyze the convergence rate of {\fontfamily{lmtt}\selectfont WAFeL} for any aggregation weights within the context of the proposed device heterogeneity model. In the analysis, the impact of each batch size is individually incorporated. Based on our findings, we introduce mismatch terms that can model the imperfections in the learning process and provide recommendations for designing the scheme accordingly.

\textit{Aggregation Cost Metric and its Optimization:} Our analysis demonstrates that aggregation weights proportional to batch sizes are optimal in the absence of estimation errors. We propose a cost metric derived from the error term in the convergence analysis, which incorporates both communication and learning factors--- specifically the MSE and mismatch terms--- to determine suitable aggregation weights that adapt to system conditions in any communication round. This supports our integrated approach to communication and learning. We also obtain an achievable bound on the error term. In addition, we suggest two solutions for optimizing the cost metric using efficient algorithms. We demonstrate that the complexity of the proposed
algorithms scales with $K^3$, where 
$K$ represents the number of devices. 

\textit{System Insights:} Our experimental findings demonstrate that the aggregation weights designed by {\fontfamily{lmtt}\selectfont
	WAFeL} exhibit significantly improved learning accuracy when compared to other over-the-air schemes, specifically about $15\%$ over the CSIT-equipped scheme \cite{huang_analog} and $30\%$ over the non-CSIT scheme \cite{cohen1}. Notably, {\fontfamily{lmtt}\selectfont
	WAFeL} achieves this performance without requiring CSIT, which makes it highly promising. Moreover, the learning performance closely approximates the ideal scenario of error-free orthogonal transmission. Finally, the superiority of the heterogeneity-aware design becomes clear when contrasted with the design that mandates a consistent batch size, set by the device with the lowest computational capacity, known as \textit{straggler} \cite{vsmith}.

\textit{Notation:} All vectors are in column form.  For vector
$\mathbf{a}$ (boldface), $a_i$
is its $i$-th component, and $\|\mathbf{a}\| = \sqrt{\mathbf{a}^\top \mathbf{a}}$
its Euclidean norm.  $\mathbf{I}_n$ is the $n \times n$ identity matrix. $\mathbf{1}$ is the all one vector. $\mathbf{a} \odot \mathbf{b}$ is the Hadamard product of vectors $\mathbf{a}$ and $\mathbf{b}$.

\section{System Model}
\begin{figure}[tb!]
\centering
\hspace{-42pt}
\includegraphics[width =4in]{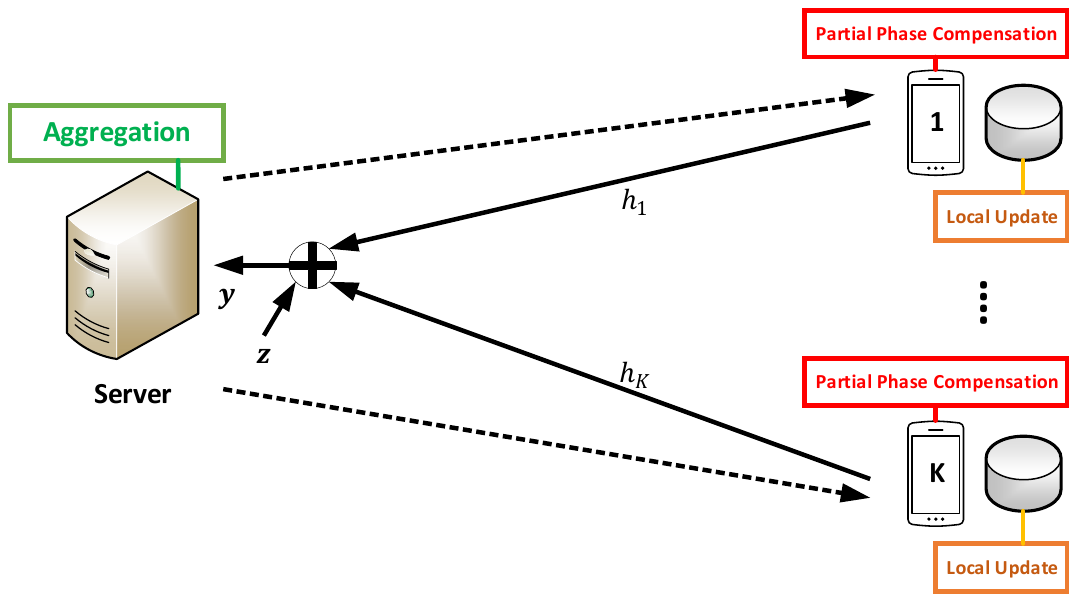}
\vspace{-237pt}
\caption{Over-the-air FL system. The phase compensation falls within the first quadrant, indicating an imperfect accuracy with a range of $[0,\frac{\pi}{2})$.}
\vspace{-5pt}
\end{figure}
\subsection{Setup}
There are $K$ devices and a single server as the basic setup for FL systems, as shown in Fig. 1. All these nodes are single-antenna units\footnote{While certain studies \cite{aircomp, wind, gunduz3, azimi_lattice, myisit, amiri22,turky} have taken into account nodes with multiple antennas, the majority of research, e.g., \cite{gunduz1, gunduz2, cao, cao1, azimi_FL, tao, huang_analog, huang_sg, huang_turning, osvaldo, wsaad, go, cohen1, cohen2, yang}, focuses on single-antenna nodes. This is due to the broad applicability and prevalent use of single-antenna nodes in wireless systems, highlighting their significance as the pivotal case for FL.}, and have frame-level
synchronization. There is no prior knowledge of local data statistics of the devices, consistent with \cite{gunduz1, gunduz2, aircomp, azimi_FL, huang_analog, huang_sg, huang_turning, osvaldo, wsaad, go, wind, gunduz3, cohen1, cohen2, yang, azimi_lattice, myisit, turky, amiri22}. The downlink channels from the server to the devices are considered error-free.\footnote{This is a valid assumption, given the high transmission power at the server and the exclusive node transmission in the downlink. It is also adopted in \cite{gunduz1, gunduz2, cao, aircomp, cao1, azimi_FL, tao, huang_analog, huang_sg, huang_turning, osvaldo, wsaad, go, wind, gunduz3, cohen1, cohen2, yang, azimi_lattice,myisit,turky,amiri22}.} The uplink channel from each device $k$ to the server at communication round $t$ is modeled by ${h}_{k,t} = |{h}_{k,t}| e^{\angle {h}_{k,t}} \in {\mathbb{C}}$, where $|{h}_{k,t}|$ is the channel gain and $\angle {h}_{k,t}$ is the channel phase. The same transmission power constraint $P$ is considered for all the devices. However, asymmetric power constraints can be
incorporated by scaling the channel coefficients appropriately. Let the entire channel vector $\mathbf{h}_t = [{h}_{1,t},\cdots,{h}_{K,t}]^\top$. The server is the only node that knows $\mathbf{h}_t$ as the CSIR \textit{after the transmission}, while each device $k$ knows only a partial estimation of its channel phase $\angle {h}_{k,t}$ with an accuracy range of $[0,\frac{\pi}{2})$. The purpose of such a \textit{quadrant phase estimation} with a wide range of inaccuracy, accounting for a quarter of the entire possible range, is to enable each device to adjust its phase so that all channels are observed as positive at the server. This ensures that the channels do not alter the sign of the transmitted data. Such partial estimation is also considered in studies like \cite{cohen1, cohen2, cohen3, cohen4, cohen5}. Hence, perfect fine synchronization is not needed, thanks to the acceptable range of uncertainty in channel phase. It is worth noting that many real-world wireless systems have CSIR and phase estimation.\footnote{The phase can be estimated using phase recovery techniques, including both data-aided and non-data-aided approaches. Extensive research is available on this topic, such as in \cite{phase}. Future research efforts can aim to even further simplify the process, with a specific focus on delivering solutions tailored to quadrant phase estimation.} Contrastingly, the majority of over-the-air FL techniques \cite{gunduz1, gunduz2, cao, cao1, azimi_FL, tao, huang_analog, huang_sg, huang_turning, osvaldo, wsaad, aircomp, go, wind} necessitate CSIT for all devices \textit{before every transmission}. This includes precise values of $|{h}_{k,t}|$ and $\angle {h}_{k,t}$ for each device $k$ and mandates perfect fine synchronization to fully counteract the channels. This places them at a complexity level distinct from our \textit{blind} approach.

Under our central assumption, we either lack knowledge of the channel gain or the information we do have contains substantial inaccuracies, thus complicating its effective use. Instead, our methodology gravitates towards the partial channel phase, providing a more reliable and consistent foundation for our proposed scheme in Section III. This is in line with other blind over-the-air FL schemes and also many conventional wireless systems, avoiding channel compensation at transmitter side \cite{gunduz3, cohen1,cohen2,yang,alphan,turky,amiri22,azimi_lattice,myisit}. This approach achieves consistent average signal energy across varying channel conditions, minimizes the dynamic range of the signal, and simplifies the complexity of device hardware. It stands out as an especially viable option for large-scale IoT applications that rely on devices with limited capabilities.\footnote{Should there be access to perfect channel gain knowledge at the devices, our scheme can be modified to adjust transmission powers accordingly, enhancing performance. However, the potential for improvement might be limited, as evidenced by the results in Figs. 8 and 9. 
Exploring this further remains an avenue for future research. In this work, our primary goal is to examine the mere potential of adjustable aggregation weights.} 
 
\vspace{0pt}
\subsection{Heterogeneity-Aware Learning Algorithm}
Device $k \in \left\{1,\cdots,K\right\}$ possesses its own local (private) dataset ${\cal D}_k$. The $i$-th sample $\xi_i$ in ${\cal D}_k$ contains a feature vector $\mathbf{x}_i$ and its ground-truth label $y_i$. The learning model is parametrized by the parameter vector $\mathbf{w} = [w_1,\cdots,w_s]^\top\in \mathbb{R}^{s\times 1}$, where $s$ is the model size. Then, the local loss function of the model vector $\mathbf{w}$ on ${\cal D}_k$ is
\begin{align}
F_k(\mathbf{w}) =  \frac{1}{D_k}\sum_{\xi_i\in {\cal D}_k}^{}\ell(\mathbf{w},\xi_i),
\end{align}
where $D_k = |{\cal D}_k|$ is the size of the dataset and the function $\ell(\mathbf{w},\xi_i)$ represents the sample-wise loss, measuring the prediction error of $\mathbf{w}$ on $\xi_i$. Following this, the global loss function applied to all distributed datasets, denoted as $\cup_{k=1}^{K} {\cal D}_k$, is 
\begin{align}
\label{lossfunction}
F(\mathbf{w}) = \frac{1}{\sum_{k=1}^{K}D_k}\sum_{k=1}^{K} D_k F_k(\mathbf{w}).
\end{align}

The goal of the training procedure is to discover an optimal parameter vector $\mathbf{w}$ that minimizes $F(\mathbf{w})$, expressed as
\begin{align}
\label{objective}
\mathbf{w}^* = \min_{\mathbf{w}} F(\mathbf{w}).
\end{align}
The typical approach of FL algorithms to solve \eqref{objective} involves two steps: local updates and aggregation, which are iterated over multiple rounds. One commonly used FL algorithm is {\fontfamily{lmtt}\selectfont
	FedAvg} \cite{mcmahan}. In the following, we propose a modified version of {\fontfamily{lmtt}\selectfont
	FedAvg} that is aware of computational heterogeneity. 

Consider a particular round $t \in \left\{0,...,T-1\right\}$, where $T$ denotes the number of rounds. In this round, each device $k$ first updates its own learning model via $\tau$ local stochastic gradient descent steps, each based on a randomly sampled mini-batch $\boldsymbol\xi_k^i$ from ${\cal D}_k$ as
\begin{align}
\label{localcompute}
\mathbf{w}_{k,t,i+1} = {\mathbf{w}_{k,t,i}}- \eta\nabla F_k(\mathbf{w}_{k,t,i}, \boldsymbol\xi_k^i), \forall i \in \left\{0,\ldots,\tau-1\right\},
\end{align}
where $\eta$ is the learning rate. Let $|\boldsymbol\xi_k^i| = B_k, \forall i$, represent the constant mini-batch size for device $k$. Each device $k$, with its distinct hardware capability determined by the computing speed of its processor (e.g., CPU or GPU) denoted as $f_k$, completes \eqref{localcompute} within the local computation time $t_k^\text{cmp} = \frac{WB_k}{f_k}$, where $W$ is the per-sample workload for local-gradient estimation \cite{cpu}. Thus, with a consistent local computation time set for all devices to meet a predetermined deadline $T_\text{p}$, i.e., $t_k^\text{cmp} = T_\text{p}, \forall k$, the batch sizes $B_k, \forall k$, reflect differences in processor computing speeds, indicating the computational heterogeneity.\footnote{
The weighted scheme discussed in the subsequent section is versatile, not confined to this heterogeneity framework, and remains novel even for the homogeneous case where $B_k = B_1, \forall k$.}
\footnote{We do not incorporate the statistical heterogeneity of devices into the design of our scheme. This approach aligns not only with other over-the-air FL research in the literature, \cite{alphan, gunduz1, gunduz2, cao, aircomp, cao1, azimi_FL, tao, huang_analog, huang_sg, huang_turning, osvaldo, wsaad, go, wind, gunduz3, cohen1, cohen2, yang, azimi_lattice,myisit,turky, amiri22}, but also with our system model, which lacks access to statistical prior knowledge about the devices. However, the effectiveness of our design on non-independent and identically distributed (non-i.i.d.) data distributions is confirmed through experiments in Section VI.} 

Then, each device $k$ uploads the local model $\mathbf{w}_{k,t} = {\mathbf{w}_{k,t,\tau}}$ to the server for aggregation. As the ideal heterogeneity-aware aggregation, the global gradient can be obtained as an average of model parameters from the devices as
\begin{align}
\label{agg}
\mathbf{w}_{\text{G},t+1} = \frac{1}{B}\sum_{k=1}^{K} B_k\mathbf{w}_{k,t},
\end{align}
where $B = \sum_{k=1}^{K}B_k$. Also, define $\mathbf{b}_\text{w} = \left[\frac{B_1}{B},\ldots,\frac{B_K}{B}\right]^\top$ as the vector containing the weights in \eqref{agg}. Next, the server broadcasts the obtained global model $\mathbf{w}_{\text{G},t+1}$ to the devices, based on which each device $k$ updates its initial state for the next round as $\mathbf{w}_{k,t+1,0} = \mathbf{w}_{\text{G},t+1}$. 

\section{{\fontfamily{lmtt}\selectfont
		WAFeL}: Weighted Over-the-Air Scheme}
The {\fontfamily{lmtt}\selectfont
	WAFeL} constructs a new form of the aggregation based on the additive nature of wireless multiple-access channels. This scheme instead of the ideal aggregation in \eqref{agg} generally has the goal to recover a weighted aggregation as
\begin{align}
\label{weighted_agg}
\mathbf{w}_{\text{G},t+1} =   \sum_{k=1}^{K}\alpha_{k,t}
\mathbf{w}_{k,t},
\end{align}
where $\alpha_{k,t}\geq 0$ is the weight of device $k$ in the aggregation at round $t$, such that $\sum_{k=1}^{K}\alpha_{k,t} = 1$. Let $\boldsymbol{\alpha}_t = [\alpha_{1,t},\ldots,\alpha_{K,t}]^\top$ be the weight vector.

Unlike the common over-the-air FL schemes in \cite{gunduz1, gunduz2, cao, cao1, azimi_FL, tao, huang_analog, huang_sg, huang_turning, osvaldo,aircomp, wsaad, go, wind}, which rely on CSIT, {\fontfamily{lmtt}\selectfont
	WAFeL} transmits blindly at a constant power. This characteristic distinguishes {\fontfamily{lmtt}\selectfont
	WAFeL} from these schemes reliant on power allocation, which enforce average and maximum power limits and inherently include device selection. In {\fontfamily{lmtt}\selectfont WAFeL}, every device contribute to learning without such restrictions, using an aggregation weight specifically designed for its communication and heterogeneity conditions. It is also worth noting that {\fontfamily{lmtt}\selectfont
	WAFeL} is not limited to the choice of learning algorithm in Subsection II.B. 

The {\fontfamily{lmtt}\selectfont
	WAFeL} comprises two principal components: the transmission scheme deployed at the devices and the aggregation scheme implemented at the server, as follows. In this section, we ignore the iteration index for simplicity of presentation.
\subsection{Transmission Scheme}
The model parameters at each
device are normalized before transmission to have a zero mean and a variance of one (unity power). Normalizing the parameters offers two benefits. First, when the parameters have zero-mean entries, the estimate obtained in Subsection III.B is unbiased. Second, when the entries have unit variance, the power of the received signal and the estimation error do not depend on the specific values of the model parameters.

The local model parameter vector at a device $k$ is normalized as ${\bar{\mathbf{w}}_{k}} = \frac{{{\mathbf{w}}_k}-\mu_k\mathbf{1}}{\sigma_k}$, where $\mathbf{1}$ is the all one vector, and $\mu_k$ and ${\sigma_k}$ denote the mean and standard deviation of the $s$ entries of the model vector given by
\begin{align}
\mu_k = \frac{1}{s}\sum_{i=1}^{s} w_{k,i},\
\sigma_k^2 = \frac{1}{s}\sum_{i=1}^{s}(w_{k,i}-\mu_k)^2,
\end{align}
where $w_{k,i}$ is the $i$-th entry of the model vector. Each device $k$ shares the two scalars $(\mu_k,\sigma_k)$ to the server without errors using an orthogonal feedback channel. This information is however negligible compared to the model parameters. 

Subsequently, after the partial phase correction, each device transmits its normalized model parameter vector by scaling it with $\sqrt{P}$, resulting in a transmission power of $P$.
\subsection{Aggregation Scheme} 
After simultaneous transmission of all the devices within a single resource block, the baseband received signal $\mathbf{y} \in \mathbb{C}^{s\times 1}$ at the server is
\begin{align}
\label{real_signal}
\mathbf{y} = \sum_{k=1}^{K}\sqrt{P}h_k \bar{\mathbf{w}}_k + \mathbf{z},
\end{align} 
where $\mathbf{z} \in \mathbb{C}^{s\times 1}$ is additive white Gaussian noise (AWGN), where each entry has variance $\sigma_\text{z}^2$. Thus, \eqref{real_signal} has the following real-valued representation
\begin{align}
\label{realsignal}
\mathbf{Y} = \sqrt{P}\mathbf{H}\bar{\mathbf{W}} + \mathbf{Z},
\end{align}
where
\begin{align}
\mathbf{Y} = \begin{bmatrix}
\mathfrak{Re}\left\{\mathbf{y^\top}\right\} \\
\mathfrak{Im}\left\{\mathbf{y^\top}\right\}
\end{bmatrix},
\end{align}
\begin{align}
\mathbf{H} = \begin{bmatrix}
\mathfrak{Re}\left\{\mathbf{h^\top}\right\} \\
\mathfrak{Im}\left\{\mathbf{h^\top}\right\}
\end{bmatrix},
\end{align}
\begin{align}
\bar{\mathbf{W}} = [\bar{\mathbf{w}}_1,\ldots,
\bar{\mathbf{w}}_K]^\top,
\end{align}
\begin{align}
\mathbf{Z} = \begin{bmatrix}
\mathfrak{Re}\left\{\mathbf{z^\top}\right\} \\
\mathfrak{Im}\left\{\mathbf{z^\top}\right\}
\end{bmatrix}.
\end{align}
\begin{figure}[tb!]
	\centering
	
	\includegraphics[width =2.8in]{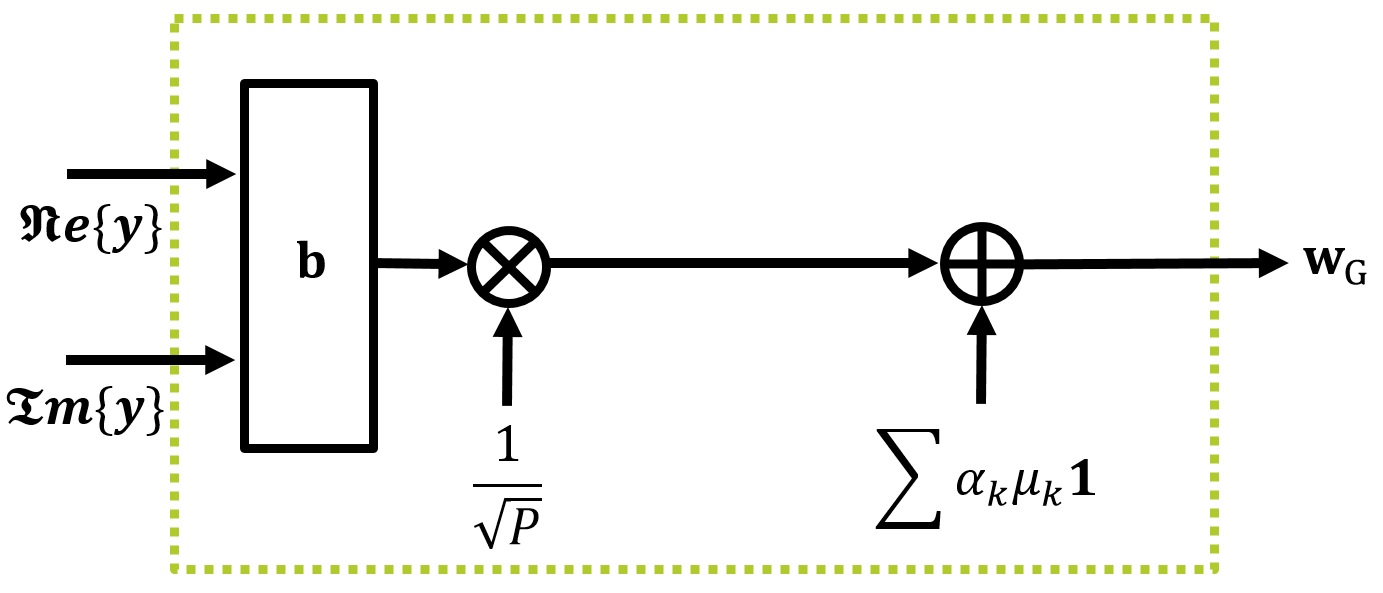}
	
	\caption{Receiver architecture at the server.}
	\vspace{-10pt}
\end{figure}
Directly from \eqref{realsignal}, the server estimates the aggregation \eqref{weighted_agg} as
\begin{align}
\hat{\mathbf{w}}_\text{G}^\top = \frac{1}{\sqrt{P}}\mathbf{b}^\top \mathbf{Y}+\sum_{k=1}^{K}\alpha_k\mu_k\mathbf{1}^\top,
\end{align}
where $\mathbf{b} \in \mathbb{R}^{2 \times 1}$ is employed as an equalization vector. The rationalization is that \eqref{weighted_agg} is recovered from $\frac{1}{\sqrt{P}}\mathbf{b}^\top \mathbf{Y}$. However, due to its zero mean resulting from the transmit normalization, the appropriate mean $\sum_{k=1}^{K}\alpha_k\mu_k\mathbf{1}^\top$ is added to ensure accurate recovery, leading to an unbiased estimation. Fig. 2 provides a schematic representation of the receiver architecture designed for this aggregation process.

We can rewrite $\hat{\mathbf{w}}_\text{G}^\top$ as 
\begin{align}
&\sum_{k=1}^{K}\alpha_k\mathbf{w}_k^\top+\frac{1}{\sqrt{P}}\mathbf{b}^\top\mathbf{Y}-\sum_{k=1}^{K}\bigl(\alpha_k\mathbf{w}_k^\top-\alpha_k\mu_k\mathbf{1}^\top\bigr)=\nonumber\\&\sum_{k=1}^{K}\alpha_k\mathbf{w}_k^\top+\mathbf{b}^\top \mathbf{H} \bar{\mathbf{W}}+ \frac{1}{\sqrt{P}}\mathbf{b}^\top \mathbf{Z}-\sum_{k=1}^{K}\alpha_k\sigma_k \bar{\mathbf{w}}_k^\top.
\end{align}
Thus, the estimation error in recovering the weighted aggregation \eqref{weighted_agg} is measured in terms of the MSE as
\begin{align}
\label{mse}
&\text{MSE}(\boldsymbol \alpha) =\nonumber\\& \mathbb{E}\left\{\left\Vert \mathbf{b}^\top \mathbf{H} \bar{\mathbf{W}}+ \frac{1}{\sqrt{P}}\mathbf{b}^\top \mathbf{Z}-(\boldsymbol\alpha\odot\boldsymbol\sigma)^\top \bar{\mathbf{W}}\right\Vert^2\right\} \nonumber\\
&=\left(\|\mathbf{b}^\top\mathbf{H}-(\boldsymbol\alpha\odot\boldsymbol\sigma)^\top\|^2+\frac{1}{\text{SNR}}\|\mathbf{b}\|^2\right){s},
\end{align}
where the independence of $\bar{\mathbf{w}}_k$ and $\bar{\mathbf{w}}_{k'}$, $\forall k \neq k'$, is assumed, similar to other over-the-air FL works \cite{power_huang, aircomp, azimi_FL, wsaad, go, wind, osvaldo, cao,cao1,azimi_lattice,myisit,gunduz3}. In \eqref{mse}, $\text{SNR} = \frac{P}{\sigma_\text{z}^2}$ is the signal-to-noise ratio. The vector $\mathbf{b}$ that minimizes the estimation error is presented in the next theorem.  
\begin{theorem}
The optimal equalization vector for a given $\boldsymbol\alpha$, minimizing the MSE in \eqref{mse}, is
\begin{align}
\mathbf{b}_\text{opt}^\top = (\boldsymbol\alpha\odot\boldsymbol\sigma)^\top \mathbf{H}^\top \left(\frac{1}{\text{SNR}}\mathbf{I}_2+\mathbf{H}\mathbf{H}^\top\right)^{-1}.
\end{align}
\end{theorem}
\begin{IEEEproof}
Expanding
\begin{align}
&\|\mathbf{b}^\top\mathbf{H}-(\boldsymbol\alpha\odot\boldsymbol\sigma)^\top\|^2+\frac{1}{\text{SNR}}\|\mathbf{b}\|^2 = \left(\mathbf{b}^\top\mathbf{H}-(\boldsymbol\alpha\odot\boldsymbol\sigma)^\top\right)\nonumber\\&\times\left(\mathbf{H}^\top\mathbf{b}-\boldsymbol\alpha\odot\boldsymbol\sigma\right)+\frac{1}{\text{SNR}}\|\mathbf{b}\|^2 = \mathbf{b}^\top\mathbf{H}\mathbf{H}^\top\mathbf{b} \nonumber\\&-2\mathbf{b}^\top\mathbf{H}(\boldsymbol\alpha\odot\boldsymbol\sigma)+(\boldsymbol\alpha\odot\boldsymbol\sigma)^\top(\boldsymbol\alpha\odot\boldsymbol\sigma)+ \frac{1}{\text{SNR}}\mathbf{b}^\top \mathbf{b},
\end{align}
and taking derivative from the result with respect to $\mathbf{b}$, we obtain
\begin{align}
2\mathbf{H}\mathbf{H}^\top \mathbf{b} - 2 \mathbf{H}(\boldsymbol\alpha\odot\boldsymbol\sigma)+\frac{2}{\text{SNR}}\mathbf{b},
\end{align}
which is equal to zero at
\begin{align}
\mathbf{b}_\text{opt}^\top =(\boldsymbol\alpha\odot\boldsymbol\sigma)^\top \mathbf{H}^\top \left(\frac{1}{\text{SNR}}\mathbf{I}_2+\mathbf{H}\mathbf{H}^\top\right)^{-1}.
\end{align}
\end{IEEEproof}
Replacing $\mathbf{b}_\text{opt}$ in \eqref{mse}, we obtain
\begin{align}
&\text{MSE}(\boldsymbol{\alpha}) = \Biggl((\boldsymbol\alpha\odot\boldsymbol\sigma)^\top\times\nonumber\\&\left( \mathbf{I}_K - \mathbf{H}^\top \left(\frac{1}{\text{SNR}}\mathbf{I}_2+\mathbf{H}\mathbf{H}^\top\right)^{-1}\mathbf{H}\right)(\boldsymbol\alpha\odot\boldsymbol\sigma)\Biggr)s,
\end{align}
which, using the matrix inversion lemma \cite{matrix_inversion}, can be written as
\begin{align}
\label{mse_final}
&\text{MSE}(\boldsymbol\alpha) = \biggl( (\boldsymbol\alpha\odot\boldsymbol\sigma)^\top\left( \mathbf{I}_K +{\text{SNR}}\mathbf{H}^\top\mathbf{H}\right)^{-1}\times\nonumber\\
&(\boldsymbol\alpha\odot\boldsymbol\sigma)\biggr)s= \biggl(\boldsymbol\alpha^\top \text{diag}(\boldsymbol\sigma)\left( \mathbf{I}_K +{\text{SNR}}\mathbf{H}^\top\mathbf{H}\right)^{-1}\times\nonumber\\
&\text{diag}(\boldsymbol\sigma)\boldsymbol\alpha\biggr)s.
\end{align}

\section{Convergence Analysis}
The analysis of {\fontfamily{lmtt}\selectfont
	WAFeL} in terms of convergence rate is presented in the following theorem. To shed light on the novelty of our analysis, it is important to emphasize that most convergence studies in this domain largely delve into scenarios characterized by equal aggregation weights, equal batch sizes, ideal communication conditions, or strongly convex loss functions. These studies navigate under an array of assumptions related to learning parameters. In contrast, our analysis boasts flexibility and is tailored to address general loss functions. This not only facilitates easy recognition of the impacts of the scheme's parameters, potential imperfections, and device heterogeneity, but also paves the way for recommending universally-applicable aggregation cost metrics, as elaborated in Section V. Notably, our analysis is based on a minimal set of assumptions commonly found in the literature, e.g., \cite{cao, cao1, azimi_FL, karimi}, as

\textbf{Assumption 1 (Lipschitz-Continuous Gradient):} The gradient of the loss function $F(\mathbf{w})$, as represented in \eqref{lossfunction}, is characterized by Lipschitz continuity with a non-negative constant $L > 0$. This implies that for any pair of model vectors $\mathbf{w}_1$ and $\mathbf{w}_2$, we have
\begin{align}
&F(\mathbf{w}_2) \leq F(\mathbf{w}_1) + \nabla F(\mathbf{w}_1)^T (\mathbf{w}_2-\mathbf{w}_1) + \frac{L}{2} \|\mathbf{w}_2 - \mathbf{w}_1\|^2,
\end{align}
\vspace{-15pt}
\begin{align}
&\|\nabla F(\mathbf{w}_2)-\nabla F(\mathbf{w}_1)\| \leq L \|\mathbf{w}_2 - \mathbf{w}_1\|.
\end{align}


\textbf{Assumption 2 (Gradient Variance Bound):} The local stochastic gradient estimate for device $k$ at $\mathbf{w}_k$, using a mini-batch $\boldsymbol \xi_k$, is an unbiased estimate
of the ground-truth gradient $\nabla F(\mathbf{w}_k)$ with bounded variance
\begin{align}
\mathbb{E}\left\{\|\nabla F_k(\mathbf{w}_{k}, \boldsymbol\xi_k) - \nabla F(\mathbf{w}_k)\|^2\right\} \leq \frac{\sigma_\text{g}^2}{B_k},
\end{align}
where $B_k = |\boldsymbol\xi_k|$ and $\sigma_\text{g}^2$ denotes the variance bound for the estimate based on a single sample. This inequality plays a key role for quantifying the computational heterogeneity via batch sizes. 
\begin{theorem}
Let $1-\frac{L^2\eta^2}{2}\tau(\tau-1)-L\eta \tau \geq 0$ and $\boldsymbol \alpha_t$ as the weight vector for each round $ t\in\left\{0,\ldots,T-1\right\}$, then the convergence rate of {\fontfamily{lmtt}\selectfont
	WAFeL} under Assumptions 1 and 2 is
\begin{align}
\label{ratebound}
&\frac{1}{T}\sum_{t=0}^{T-1}\mathbb{E}\left\{\|\nabla F( {\mathbf{w}}_{\text{G},t})\|^2\right\} \leq  \frac{2\left(F({\mathbf{w}}_{\text{G},0})-F^*\right)}{\eta \tau T}\nonumber\\
&+\frac{L}{\eta \tau T}\sum_{t=0}^{T-1}{\cal I}_t(\boldsymbol \alpha_t),
\end{align}
where $F^* = F(\mathbf{w}^*)$ comes from the problem \eqref{objective} as the optimal loss value, and
\begin{align}
\label{mainL}
{\cal I}_t(\boldsymbol \alpha_t) &= {L\eta^3}\frac{\tau(\tau-1)}{2}{\sigma_\text{g}^2} \boldsymbol \alpha_t^\top \mathbf{b}_\text{s}+{ \eta^2}{\sigma_\text{g}^2}\tau \boldsymbol \alpha_t^\top \text{diag}\left\{\mathbf{b}_\text{s}\right\} \boldsymbol \alpha_t\nonumber\\&+ {\text{MSE}_t(\boldsymbol \alpha_t)},
\end{align}
where $\text{MSE}_t$ is given in \eqref{mse_final} and the vector $\mathbf{b}_\text{s} = [\frac{1}{B_1},\ldots,\frac{1}{B_K}]^\top$ is formed from the inverse of batch sizes of all the devices.
\end{theorem}
\begin{IEEEproof}
	See Appendix A.
\end{IEEEproof}
\begin{remark}
The convergence rate in \eqref{ratebound} incorporates communication and learning factors such as the estimation error measured by $\text{MSE}$ including wireless channels and SNR, the aggregation weights, the batch sizes, the learning rate, the number of local steps, and the gradient variance bound. This emphasizes the integrated approach of {\fontfamily{lmtt}\selectfont WAFeL}, blending communication and learning within its design.
\end{remark}
\begin{remark}
Increasing $\eta$ or $\tau$ directly reduces the first term in the convergence rate but might raise the second term, known as the error term. On the other hand, other factors exclusively affect the error term. Raising batch sizes, or a reduction in $\text{MSE}$ or $\sigma_\text{g}^2$, diminishes the error term and consequently improves performance.	
\end{remark}

From Theorem 2, the convergence rate of the proposed version of {\fontfamily{lmtt}\selectfont FedAvg} in Subsection II.B--- under error-free transmission and aggregation \eqref{agg}--- serves as a specific case of {\fontfamily{lmtt}\selectfont WAFeL} and is presented below.
\begin{corollary}
Let $1-\frac{L^2\eta^2}{2}\tau(\tau-1)-L\eta \tau \geq 0$ and $\boldsymbol \alpha_t = \mathbf{b}_\text{w}$ for each round $ t\in\left\{0,\ldots,T-1\right\}$, then in the case of error-free transmission, the convergence rate under Assumptions 1 and 2 is
\begin{align}
&\frac{1}{T}\sum_{t=0}^{T-1}\mathbb{E}\left\{\|\nabla F( {\mathbf{w}}_{\text{G},t})\|^2\right\} \leq  \frac{2\left(F({\mathbf{w}}_{\text{G},0})-F^*\right)}{\eta \tau T}\nonumber\\&+\frac{L}{\eta \tau T}\sum_{t=0}^{T-1}{\cal I},
\end{align}
where
\begin{align}
\label{mainI}
{\cal I} = {L\eta^3}\frac{\tau(\tau-1)}{2}\frac{{\sigma_\text{g}^2} }{B}K+{ \eta^2}\tau \frac{\sigma_\text{g}^2}{B}.
\end{align}
\end{corollary}
\begin{remark}
The effect of errors in {\fontfamily{lmtt}\selectfont WAFeL} is an increase in the ${\cal I}$ term by ${L\eta^3}\frac{\tau(\tau-1)}{2}{\sigma_\text{g}^2} \left(\boldsymbol \alpha_t^\top \mathbf{b}_\text{s}- \frac{K}{B}\right)+{ \eta^2}{\sigma_\text{g}^2}\tau \Bigl(\boldsymbol \alpha_t^\top \text{diag}\left\{\mathbf{b}_\text{s}\right\} \boldsymbol \alpha_t-\frac{1}{B}\Bigr) + {\text{MSE}_t(\boldsymbol \alpha_t)}$ for each round $t$. This increase is due not only to the MSE term resulting from the estimation error, but also from the mismatch between the weight vector $\boldsymbol\alpha_t$ and $\mathbf{b}_\text{w}$, the latter being used for the ideal aggregation in \eqref{agg}.	
\end{remark}
\begin{remark}
The mismatch terms $\boldsymbol \alpha_t^\top \mathbf{b}_\text{s}- \frac{K}{B}$ and $\boldsymbol \alpha_t^\top \text{diag}\left\{\mathbf{b}_\text{s}\right\} \boldsymbol \alpha_t-\frac{1}{B}$ are related to the learning facet of {\fontfamily{lmtt}\selectfont
	WAFeL} and the MSE term ${\text{MSE}_t(\boldsymbol \alpha_t)}$ is due to the communication facet of {\fontfamily{lmtt}\selectfont
	WAFeL}.
\end{remark}
\begin{remark}
The mismatch term $\boldsymbol \alpha_t^\top \text{diag}\left\{\mathbf{b}_\text{s}\right\} \boldsymbol \alpha_t-\frac{1}{B}$ is scaled by ${ \eta^2}\frac{\sigma_\text{g}^2}{B}\tau$. This indicates that the effect of the mismatch is reinforced by an increase in the local training parameters $\eta$ and $\tau$.
\end{remark}
\begin{remark}
	The scaling factor of the mismatch term $\boldsymbol \alpha_t^\top \mathbf{b}_\text{s}- \frac{K}{B}$ relative to the scaling factor of the mismatch term $\boldsymbol \alpha_t^\top \text{diag}\left\{\mathbf{b}_\text{s}\right\} \boldsymbol \alpha_t-\frac{1}{B}$ is equal to $L\eta \frac{\tau-1}{2}$. Consequently, when $\eta \ll \frac{2}{L(\tau-1)}$, the impact of the first term on convergence is negligible compared to the latter.
\end{remark}
\begin{remark}
In the case of equal batch sizes $B_k = B_{1}, \forall k>1$, i.e., $\mathbf{b}_\text{w} = \frac{1}{K} \mathbf{1}$ and $\mathbf{b}_\text{s} = \frac{1}{B_1} \mathbf{1}$, the mismatch term $\boldsymbol \alpha_t^\top \mathbf{b}_\text{s}- \frac{K}{B} = \frac{1}{B_1}\boldsymbol \alpha_t^\top \mathbf{1} - \frac{1}{B_1} = 0$ disappears and the mismatch term $\boldsymbol \alpha_t^\top \text{diag}\left\{\mathbf{b}_\text{s}\right\} \boldsymbol \alpha_t-\frac{1}{B} = \frac{1}{B_1} \left(\|\boldsymbol \alpha_t\|^2 - \frac{1}{K}\right)$.
\end{remark}
\begin{remark}
When \( \tau = 1 \), {\fontfamily{lmtt}\selectfont
	FedSGD} \cite{mcmahan} is the special error-free case of {\fontfamily{lmtt}\selectfont
	WAFeL}. In this case, the mismatch term \( \boldsymbol \alpha_t^\top \mathbf{b}_\text{s} - \frac{K}{B} \) scales to zero, rendering it without impact.
\end{remark}
\begin{remark}
When there is no estimation error, the minimum of the error term ${\cal I}_t(\boldsymbol \alpha_t)$ happens at ${{\boldsymbol\alpha}_t} = \mathbf{b}_\text{w}$, as both mismatch terms become zero. That is why conventionally the ideal aggregation in the the primary {\fontfamily{lmtt}\selectfont
	FedAvg} in \cite{mcmahan} and the literature aligns with \eqref{agg}, meaning it is proportional to dataset sizes. It is demonstrated here that in the presence of estimation error, the optimal aggregation that minimizes the term ${\cal I}_t(\boldsymbol \alpha_t)$ deviates from the ideal aggregation. This is further discussed in Section V.
\end{remark}

\section{Aggregation Weight Selection}
To select the weight vectors $\boldsymbol\alpha_t, \forall t \in \left\{0,\ldots,T-1\right\}$, we minimize the error term $\sum_{t=0}^{T-1}{\cal I}_t(\boldsymbol \alpha_t)$ in the convergence rate in Theorem 2, denoted as the aggregation cost metric. It is inspired by \textit{Remark 9}. Error terms specific to convergence rates have also been widely used for optimizations in other FL research works, e.g. \cite{error1,error2}. Given the summation of ${\cal I}_t(\boldsymbol \alpha_t)$ and the independence of the condition $\mathbf{1}^\top\boldsymbol \alpha_t = 1$ across different values of $t$, this minimization can be equally decomposed into a
series of subproblems each for one round $t$ as follows. 
\begin{align}
\label{main_obj}
&{\boldsymbol\alpha_t} = \arg\min_{\boldsymbol\alpha\backslash \left\{\mathbf{0}\right\},\ \mathbf{1}^\top \boldsymbol\alpha = 1} {\cal I}_t(\boldsymbol\alpha),
\end{align}
where from \eqref{mainL}
\begin{align}
&{\cal I}_t(\boldsymbol\alpha) = {L\eta^3}\frac{\tau(\tau-1)}{2}{\sigma_\text{g}^2} \boldsymbol \alpha^\top \mathbf{b}_\text{s}+{ \eta^2}{\sigma_\text{g}^2}\tau \boldsymbol \alpha^\top \text{diag}\left\{\mathbf{b}_\text{s}\right\} \boldsymbol \alpha \nonumber\\&+ s\boldsymbol\alpha^\top \text{diag}(\boldsymbol \sigma_t)\left( \mathbf{I}_K +{\text{SNR}}\mathbf{H}_t^\top\mathbf{H}_t\right)^{-1}\text{diag}(\boldsymbol \sigma_t)\boldsymbol \alpha \nonumber\\&={L\eta^3}\frac{\tau(\tau-1)}{2}{\sigma_\text{g}^2} \boldsymbol \alpha^\top  \mathbf{b}_\text{s}+\boldsymbol\alpha^\top
\biggl(s\text{diag}(\boldsymbol\sigma_t)\nonumber\\&\left( \mathbf{I}_K +{\text{SNR}}\mathbf{H}_t^\top\mathbf{H}_t\right)^{-1}\text{diag}(\boldsymbol\sigma_t)+{ \eta^2}{\sigma_\text{g}^2}\tau\text{diag}\left\{\mathbf{b}_\text{s}\right\}\biggr)\boldsymbol\alpha. \nonumber
\end{align}
It can be written as
\begin{align}
\label{alak_obj}
&\boldsymbol\alpha_t = \arg\min_{\boldsymbol\alpha\backslash \left\{\mathbf{0}\right\}} {\boldsymbol\alpha^\top\mathbf{G}_t\boldsymbol\alpha}+\boldsymbol\alpha^\top \mathbf{c},
\end{align}
subject to
\begin{align}
\mathbf{1}^\top \boldsymbol\alpha = 1,\nonumber
\end{align}
where 
\begin{align}
\mathbf{G}_t &= s\text{diag}(\boldsymbol\sigma_t)\left( \mathbf{I}_K +{\text{SNR}}\mathbf{H}_t^\top\mathbf{H}_t\right)^{-1}\text{diag}(\boldsymbol\sigma_t)\nonumber\\&+{ \eta^2}{\sigma_\text{g}^2}\tau\text{diag}\left\{\mathbf{b}_\text{s}\right\},
\end{align}
and
\begin{align}
\mathbf{c} = {L\eta^3}\frac{\tau(\tau-1)}{2}{\sigma_\text{g}^2}  \mathbf{b}_\text{s}.
\end{align}
The problem \eqref{alak_obj} exhibits convexity. The solution of this problem is presented in the following theorem.
\begin{theorem}
The optimized weight vector as the solution of \eqref{alak_obj} is
\begin{align}
\label{optimal_alpha}
\boldsymbol{\alpha}_t = \mathbf{G}_t^{-1} \left(\frac{1+\frac{1}{2}\mathbf{1}^\top \mathbf{G}_t^{-1}\mathbf{c}}{\mathbf{1}^\top \mathbf{G}_t^{-1}\mathbf{1}}\mathbf{1}-\frac{1}{2}\mathbf{c}\right).
\end{align}	
\end{theorem}
\begin{IEEEproof}
We use the Karush-Kuhn-Tucker (KKT) conditions \cite{boyd} to solve \eqref{alak_obj}. The corresponding Lagrangian function is as follows
\begin{align}
{\cal L}(\xi, \boldsymbol\alpha) = \boldsymbol{\alpha}^\top\mathbf{G}_t\boldsymbol{\alpha} +\boldsymbol{\alpha}^\top \mathbf{c}+ \xi \boldsymbol{\alpha}^\top \mathbf{1},
\end{align}	
which, after taking derivative, leads to
\begin{align}
2\mathbf{G}_t \boldsymbol{\alpha}+\mathbf{c}+\xi\mathbf{1} = \mathbf{0},
\end{align}
and then
\begin{align}
\label{alp_proof}
\boldsymbol{\alpha} = -\frac{1}{2}\mathbf{G}_t^{-1}\left(\xi\mathbf{1}+\mathbf{c}\right).
\end{align}
To meet the condition $\mathbf{1}^\top \boldsymbol\alpha = 1$, $\xi$ is obtained as
\begin{align}
\xi = \frac{-2-\mathbf{1}^\top \mathbf{G}_t^{-1}\mathbf{c}}{\mathbf{1}^\top \mathbf{G}_t^{-1}\mathbf{1}} < 0,
\end{align}
which, when replaced in \eqref{alp_proof}, completes the proof. Given that $\mathbf{G}_t$ is a positive definite matrix with all non-negative entries due to the phase compensation, its inverse, $\mathbf{G}_t^{-1}$, is also positive definite with all non-negative entries. As a result, every entry of $\mathbf{G}_t^{-1}\mathbf{1}$ is positive. Given the relative small values in $\mathbf{c}$ attributed to the inclusion of a higher order of $\eta$, this implies that all entries of the optimal solution $\boldsymbol \alpha_t$ are also positive, which is essential for the aggregation as defined for \eqref{weighted_agg}. Indeed, the purpose of phase compensation is to guarantee that minimizing the objective ${\boldsymbol\alpha^\top\mathbf{G}_t\boldsymbol\alpha}+\boldsymbol\alpha^\top \mathbf{c}$ over weight vectors with solely nonnegative elements is equivalent to minimizing the objective ${\boldsymbol\alpha^\top\mathbf{G}_t\boldsymbol\alpha}+\boldsymbol\alpha^\top \mathbf{c}$ across all weight vectors, whether positive or negative. Therefore, the lowest value of the objective is achieved by a weight vector without any constraints, which is also appropriate for aggregation. In the absence of phase compensation, the solution might not necessarily consist of all positive elements, rendering it unsuitable for aggregation. An alternative is to seek another weight vector with entirely positive entries to minimize the objective. However, this approach is detrimental to performance, especially with a single-antenna server, a primary assumption in this study.

\end{IEEEproof}
After mathematical simplifications, the optimal $\boldsymbol{\alpha}_t$ results in the following achievable bound on the error term.
\begin{align}
\label{uselessmse}
&{\cal I}_t(\boldsymbol\alpha_t)= \frac{\left(1+\frac{1}{2}\mathbf{1}^\top \mathbf{G}_t^{-1}\mathbf{c}\right)^2}{\mathbf{1}^\top\mathbf{G}_t^{-1}\mathbf{1}} - \frac{1}{4} \mathbf{c}^\top\mathbf{G}_t^{-1}\mathbf{c},
\end{align}
which, under the special case of $\tau = 1$, is simplified to
\begin{align}
\label{specialcase}
&{\cal I}_t(\boldsymbol\alpha_t)= \frac{1}{\mathbf{1}^\top\mathbf{G}_t^{-1}\mathbf{1}} \nonumber\\&= \frac{1}{\sum_{k=1}^{K} \left(s\sigma_{k,t}^{2}\left(1+\text{SNR} v_{k,t}^2\right)^{-1}+\frac{{ \eta^2}{\sigma_\text{g}^2}}{B_k}\right)^{-1}},
\end{align}
where $v_{k,t}$ denotes the $k$-th singular value of matrix $\mathbf{H}_t$.
\begin{remark}
	If the matrix $\mathbf{H}_t$, which characterizes the channel conditions at round $t$, has higher singular values, it results in a decreased ${\cal I}_t$ and subsequently superior learning performance.
\end{remark}

Given that the precise determination of the gradient variance bound $\sigma_\text{g}^2$ and the Lipschitz constant $L$ necessitates an understanding of gradient data statistics--- which is unavailable in many applications--- the optimal evaluation of $\boldsymbol \alpha_t$ in \eqref{optimal_alpha} becomes impractical. Therefore, in pursuit of our universality goal, we introduce two alternative algorithmic solutions that do not rely on this specific knowledge.\footnote{While most FL studies do not assume such knowledge like our approach, there are research works that assume its availability \cite{go, cao, cao1, tao}. However, the result in Theorem 3 is insightful, and valuable when the knowledge is available. Also, \eqref{uselessmse} and \eqref{specialcase} can be viewed as achievable bounds for the error term in the convergence rate, providing valuable analytical utility.}
\vspace{0pt}
\subsection{MSE minimization} We consider the MSE of the estimation error as the main part of the aggregation cost metric to be minimized, while the mismatch term $\boldsymbol \alpha^\top \text{diag}\left\{\mathbf{b}_\text{s}\right\} \boldsymbol \alpha-\frac{1}{B}$ is bounded by a threshold. From \textit{Remark 4}, this ensures joint learning and communication design. However, inspired by \textit{Remarks 6, 7, and 8} and for tractability, we overlook the mismatch term $\boldsymbol \alpha^\top \mathbf{b}_\text{s}- \frac{K}{B}$. Another justification for this decision is that the minimum for both mismatch terms occurs when $\boldsymbol{\alpha} = \mathbf{b}_\text{w}$. Therefore, minimizing $\boldsymbol \alpha^\top \text{diag}\left\{\mathbf{b}_\text{s}\right\} \boldsymbol \alpha-\frac{1}{B}$ can effectively diminish $\boldsymbol \alpha^\top \mathbf{b}_\text{s}- \frac{K}{B}$. It also aligns better with the common trend of using the L2 norm in optimization problems compared to the L1 norm \cite{boyd}.\footnote{It is worth noting that in other over-the-air FL schemes, like those in \cite{wsaad, aircomp, go, wind}, providing a metric that encompasses all possible factors has not been achievable. They heuristically use the inverse of the number of participating devices as their device selection (aggregation cost) metric.} We can frame the problem as follows.
\begin{align}
\label{practical_obj1}
&\boldsymbol\alpha_t =\nonumber\\& \arg\min_{\boldsymbol\alpha\backslash \left\{\mathbf{0}\right\}} {\boldsymbol\alpha^\top\text{diag}(\boldsymbol\sigma_t)\left( \mathbf{I}_K +{\text{SNR}}\mathbf{H}_t^\top\mathbf{H}_t\right)^{-1}\text{diag}(\boldsymbol\sigma_t)\boldsymbol\alpha}, 
\end{align}
subject to
\begin{align}
&\boldsymbol \alpha^\top \text{diag}\left\{\mathbf{b}_\text{s}\right\} \boldsymbol \alpha \leq \text{th}_1,\nonumber\\&\mathbf{1}^\top \boldsymbol{\alpha} = 1.\nonumber
\end{align}
The problem \eqref{practical_obj1} is convex. Using the dual Lagrangian method \cite{boyd}, it can be transformed to
\begin{align}
\label{dual1}
&\max_{\lambda>0}\min_{\alpha\backslash \left\{\mathbf{0}\right\}}\biggl\{{\cal L}_1(\lambda, \boldsymbol{\alpha}) = \boldsymbol\alpha^\top\text{diag}(\boldsymbol\sigma_t)\left( \mathbf{I}_K +{\text{SNR}}\mathbf{H}_t^\top\mathbf{H}_t\right)^{-1}\nonumber\\
&\text{diag}(\boldsymbol\sigma_t)\boldsymbol\alpha + \lambda \boldsymbol \alpha^\top \text{diag}\left\{\mathbf{b}_\text{s}\right\} \boldsymbol \alpha = \boldsymbol\alpha^\top\times\nonumber\\
&\left(\text{diag}(\boldsymbol\sigma_t)\left( \mathbf{I}_K +{\text{SNR}}\mathbf{H}_t^\top\mathbf{H}_t\right)^{-1}\text{diag}(\boldsymbol\sigma_t)+ \lambda \text{diag}\left\{\mathbf{b}_\text{s}\right\}\right)\boldsymbol\alpha \biggr\},
\end{align}
subject to
\begin{align}
\mathbf{1}^\top \boldsymbol{\alpha} = 1.\nonumber
\end{align}
In \eqref{dual1}, the following problem
\begin{align}
\label{g1}
&g_1(\lambda) = \min_{\alpha\backslash \left\{\mathbf{0}\right\}} \boldsymbol\alpha^\top\times \nonumber\\&\left(\text{diag}(\boldsymbol\sigma_t)\left( \mathbf{I}_K +{\text{SNR}}\mathbf{H}_t^\top\mathbf{H}_t\right)^{-1}\text{diag}(\boldsymbol\sigma_t)+ \lambda \text{diag}\left\{\mathbf{b}_\text{s}\right\}\right)\boldsymbol\alpha,
\end{align}
subject to
\begin{align}
\mathbf{1}^\top \boldsymbol{\alpha} = 1,\nonumber
\end{align}
has a solution that is analogous to the one found in Theorem 3, replacing $\mathbf{c} = \mathbf{0}$ and $\mathbf{G}_{t}$ with $\mathbf{G}_{1,t}(\lambda)$, as
\begin{align}
\label{alp_it1}
\boldsymbol{\alpha} = \frac{\mathbf{G}_{1,t}(\lambda)^{-1}\mathbf{1}}{\mathbf{1}^\top\mathbf{G}_{1,t}(\lambda)^{-1}\mathbf{1}},
\end{align}
where
\begin{align}
\mathbf{G}_{1,t}(\lambda) &= \text{diag}(\boldsymbol\sigma_t)\left( \mathbf{I}_K +{\text{SNR}}\mathbf{H}_t^\top\mathbf{H}_t\right)^{-1}\text{diag}(\boldsymbol\sigma_t) \nonumber\\&+ \lambda \text{diag}\left\{\mathbf{b}_\text{s}\right\}.
\end{align}
Now, the remaining work is to maximize $g_1(\lambda)$ with respect to nonnegative $\lambda$. We use the subgradient method \cite{boyd} to find a solution. The subgradient direction of the function
$g_1(\lambda)$ is determined as $\boldsymbol \alpha^\top \text{diag}\left\{\mathbf{b}_\text{s}\right\} \boldsymbol \alpha - \text{th}_1$,
where $\boldsymbol{\alpha}$ is given in \eqref{alp_it1} with the fixed $\lambda$. Subsequently, the solution for problem \eqref{practical_obj1} can be searched in an iterative fashion summarized in Algorithm 1.

\subsection{Mismatch minimization} Here, the main part of the aggregation cost metric for minimization is the mismatch, under the condition that the MSE is confined within a certain threshold. This can be expressed as follows.
\begin{align}
\label{practical_obj2}
\boldsymbol\alpha_t = \arg\min_{\boldsymbol\alpha\backslash \left\{\mathbf{0}\right\}} \boldsymbol \alpha^\top \text{diag}\left\{\mathbf{b}_\text{s}\right\} \boldsymbol \alpha, 
\end{align}
subject to
\begin{align}
&{\boldsymbol\alpha^\top\text{diag}(\boldsymbol\sigma_t)\left( \mathbf{I}_K +{\text{SNR}}\mathbf{H}_t^\top\mathbf{H}_t\right)^{-1}\text{diag}(\boldsymbol\sigma_t)\boldsymbol\alpha} \leq \text{th}_2.\nonumber \\&\mathbf{1}^\top \boldsymbol{\alpha} = 1.\nonumber
\end{align}
Similar to the proposed approach for \eqref{practical_obj1}, we consider the dual problem
\begin{align}
&\max_{\lambda>0}\min_{\alpha\backslash \left\{\mathbf{0}\right\}}\biggl\{{\cal L}_2(\lambda, \boldsymbol{\alpha}) =\boldsymbol \alpha^\top \text{diag}\left\{\mathbf{b}_\text{s}\right\} \boldsymbol \alpha+\nonumber\\
&\lambda {\boldsymbol\alpha^\top\text{diag}(\boldsymbol\sigma_t)\left( \mathbf{I}_K +{\text{SNR}}\mathbf{H}_t^\top\mathbf{H}_t\right)^{-1}\text{diag}(\boldsymbol\sigma_t)\boldsymbol\alpha}  = \boldsymbol\alpha^\top\times\nonumber\\&\left( \text{diag}\left\{\mathbf{b}_\text{s}\right\}+\lambda\text{diag}(\boldsymbol\sigma_t)\left( \mathbf{I}_K +{\text{SNR}}\mathbf{H}_t^\top\mathbf{H}_t\right)^{-1}\text{diag}(\boldsymbol\sigma_t)\right)\boldsymbol\alpha\biggr\},
\end{align}
subject to
\begin{align}
\mathbf{1}^\top \boldsymbol{\alpha} = 1,\nonumber
\end{align}
where the solution of the problem
\begin{align}
\label{g2}
&g_2(\lambda) = \min_{\alpha\backslash \left\{\mathbf{0}\right\}} \boldsymbol\alpha^\top\times \nonumber\\
&\left(\text{diag}\left\{\mathbf{b}_\text{s}\right\}+\lambda\text{diag}(\boldsymbol\sigma_t)\left( \mathbf{I}_K +{\text{SNR}}\mathbf{H}_t^\top\mathbf{H}_t\right)^{-1}\text{diag}(\boldsymbol\sigma_t)\right)\boldsymbol\alpha.
\end{align}
subject to
\begin{align}
\mathbf{1}^\top \boldsymbol{\alpha} = 1,\nonumber
\end{align}
is
\begin{align}
\label{alp_it2}
\boldsymbol{\alpha} = \frac{\mathbf{G}_{2,t}(\lambda)^{-1}\mathbf{1}}{\mathbf{1}^\top\mathbf{G}_{2,t}(\lambda)^{-1}\mathbf{1}},
\end{align}
with
\begin{align}
\mathbf{G}_{2,t}(\lambda) &= \text{diag}\left\{\mathbf{b}_\text{s}\right\}+\nonumber\\&\lambda\text{diag}(\boldsymbol\sigma_t)\left( \mathbf{I}_K +{\text{SNR}}\mathbf{H}_t^\top\mathbf{H}_t\right)^{-1}\text{diag}(\boldsymbol\sigma_t).
\end{align}
Hence, similar to Algorithm 1, the solution of problem \eqref{practical_obj2} can be obtained by the iterative Algorithm 2.

Before we conclude this section, it is worth noting that neither algorithm generally holds precedence over the other. The choice between them depends on their respective abilities to determine an appropriate $\text{th}_1$ or $\text{th}_2$, and whether emphasis is placed on the communication aspect of the scheme, MSE, or its learning aspect, mismatch. These are counted as hyperparameters, obtained by fine-tuning, though finding their optimum is not the focus of this study. Furthermore, if CSIT is available, one can use transmission powers and their optimization to further decrease the MSE. Finally, extending $\mathbf{H}_t$ appropriately allows for straightforward extension of the scheme to the multiple-antenna server case.

\begin{algorithm}
	\caption{Aggregation weight selection from \eqref{practical_obj1}}
\begin{algorithmic}
			\vspace{0pt}
			\State Initialize $\lambda^{(0)}$ and $\boldsymbol \alpha^{(0)} = \frac{\text{diag}(\bar{\boldsymbol\sigma}_t)\left( \mathbf{I}_K +{\text{SNR}}\mathbf{H}_t^\top\mathbf{H}_t\right)\text{diag}(\bar{\boldsymbol\sigma}_t)\mathbf{1}}{\mathbf{1}^\top\text{diag}(\bar{\boldsymbol\sigma}_t)\left( \mathbf{I}_K +{\text{SNR}}\mathbf{H}_t^\top\mathbf{H}_t\right)\text{diag}(\bar{\boldsymbol\sigma}_t)\mathbf{1}}$, {\small{where $\bar{\boldsymbol\sigma}_t = \left[1/\sigma_{1,t}, \cdots, 1/\sigma_{K,t}\right]$.}}
			
			\State Iterate  
			\State \hspace{20pt}Update $\boldsymbol \alpha^{(j)}$ as in \eqref{alp_it1} with $\lambda^{(j-1)}$.
			\State \hspace{20pt}Update $\lambda^{(j)}$ as {\small{$\lambda^{(j)} = \lambda^{(j-1)}+t\left(\boldsymbol {\alpha^{(j)}}^\top \text{diag}\left\{\mathbf{b}_\text{s}\right\} \boldsymbol \alpha^{(j)} - \text{th}_1\right)$}}.
			\State Until {\small{$\Big|\lambda \left(\boldsymbol {\alpha}^\top \text{diag}\left\{\mathbf{b}_\text{s}\right\} \boldsymbol \alpha - \text{th}_1\right)\Big|\leq \epsilon$}}.
	\end{algorithmic}
\end{algorithm}

\begin{algorithm}
	\caption{Aggregation weight selection from \eqref{practical_obj2}}
	\begin{algorithmic}
		\vspace{0pt}
		\State Initialize $\lambda^{(0)}$ and $\boldsymbol \alpha^{(0)} = \mathbf{b}_\text{w}$
		\State Iterate  
		\State \hspace{20pt}Update $\boldsymbol \alpha^{(j)}$ as in \eqref{alp_it2} with $\lambda^{(j-1)}$.
		\State \hspace{20pt}Update $\lambda^{(j)}$ as {\small{$\lambda^{(j)} = \lambda^{(j-1)}+t\left({{\boldsymbol\alpha^{(j)}}^\top\text{diag}(\boldsymbol\sigma_t)\left( \mathbf{I}_K +{\text{SNR}}\mathbf{H}_t^\top\mathbf{H}_t\right)^{-1}\text{diag}(\boldsymbol\sigma_t)\boldsymbol\alpha^{(j)}}  - \text{th}_2\right)$}}.
		\State Until \small{$\Big|\lambda \left({\boldsymbol\alpha^\top\text{diag}(\boldsymbol\sigma_t)\left( \mathbf{I}_K +{\text{SNR}}\mathbf{H}_t^\top\mathbf{H}_t\right)^{-1}\text{diag}(\boldsymbol\sigma_t)\boldsymbol\alpha}  - \text{th}_2\right)\Big|\leq \epsilon$}.
	\end{algorithmic}
\end{algorithm}
\vspace{-10pt}
\subsection{Complexity Analysis}
The computational complexity involved in calculating the inverse of a $K \times K$ matrix, as required for either \eqref{alp_it1} or \eqref{alp_it2}, is on the order of $\mathcal{O}(K^3)$. Thus, for a maximum number of iterations $n_\text{max}$ in Algorithms 1 and 2, the algorithms have the complexity order of $\mathcal{O}(n_\text{max}K^3)$.
\section{Experimental Results}
The learning task is the classification
on the standard MNIST and CIFAR-10 datasets with the parameter
values given in Table 1, unless otherwise stated.\footnote{The parameters $\eta$ and $\epsilon$ are obtained by fine-tuning.} 
	The classifier models for MNIST and CIFAR-10 are built on a convolutional neural network (CNN), differing in layer configuration. For MNIST, the model features two $3 \times 3$ convolution layers with ReLU activation, the first layer having 32 channels and the second 64 channels, each followed by a $2 \times 2$ max pooling layer. In contrast, the CIFAR-10 model expands to four $3 \times 3$ convolution layers with ReLU activation, where the initial two layers are configured with 32 channels and the subsequent two with 64 channels, with a $2 \times 2$ max pooling layer after every second layer. Both architectures share a common structure thereafter: a fully connected layer with 128 units and ReLU activation, culminating in a softmax output layer for classification \cite{azimi_FL}. We take into account both independent and identically distributed (i.i.d.) and non-i.i.d. distribution of dataset samples across the devices. In the non-i.i.d. scenario, each device holds samples from only two classes, and the quantity of samples varies among devices. To account for computational heterogeneity, each device $k$ possesses distinct processor computing speed $f_k$, leading it to select a fixed batch size $B_k$ from the range between 20 and 60 based on the deadline $T_\text{p}$, with the relationship $\frac{T_\text{p}}{W}f_k = B_k \in [20, 60], \forall k$. Through which, the fastest device exhibits processing power three times greater than that of the slowest device. Performance is gauged in terms of learning accuracy relative to the test dataset and training loss across global iterations (denoted as $t$), aligning with standard metrics used in existing literature, for instance, \cite{gunduz1, gunduz2, go, wind, yang, gunduz3, tao, huang_sg, huang_analog,osvaldo, azimi_FL,aircomp,wsaad,turky,amiri22,azimi_lattice,myisit,cao,cao1}. Each
performance result is evaluated as the average of
20 realization samples to account for Gaussian channel distribution, i.e., the channel gain with Rayleigh fading $\sim \exp(1)$ and the channel phase (after the partial phase compensation) with uniform distribution $\sim {\cal{U}}(0,\frac{\pi}{2})$.
\begin{table}
	\centering 
	\caption{Parameter Values}
	\vspace{-5pt} 
	\resizebox{0.5\textwidth}{!}{ 
		\begin{tabular}{| l | l | l | l | l | l | l | l | l |}
			\hline
			$K$ & $\text{SNR}$ & $\tau$ & $\eta$ & $B_k, \forall k$ & $\epsilon$ & $W$ & $T_\text{p}$ & $f_k$ \\ \hline
			30 & 10 & 3 & 0.01 & $[20,60]$ & $0.0001$ & $1.09 \times 10^{6}$ & $0.0218\ \text{s}$ & $[1, 3] \times 10^9\ \text{Hz}$ \\
			\hline
		\end{tabular}
	}
	\vspace{-7pt} 
\end{table}

\begin{figure}[tb!]
	\vspace{0pt}
	\centering
	\includegraphics[width =2.65in]{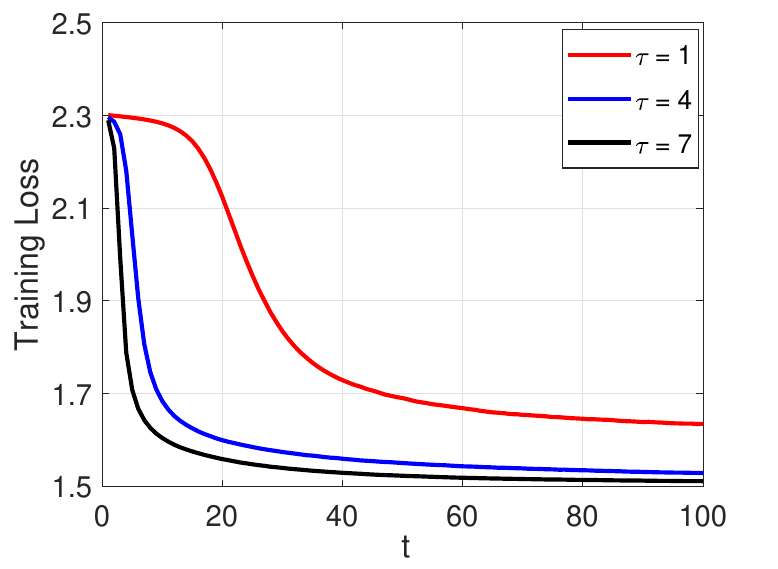} 
	\vspace{-5pt}
	\caption{Algorithm 2, MNIST, i.i.d. case.}
	\label{fig:tau}
	\vspace{-5pt}
\end{figure}

\begin{figure}[tb!]
	\vspace{0pt}
	\centering
	\includegraphics[width =2.65in]{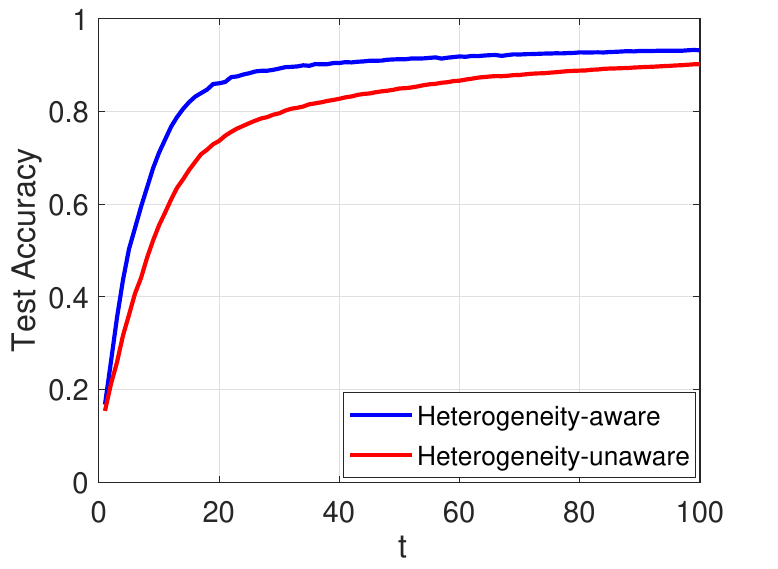}
	\vspace{-5pt}
	\caption{Algorithm 2, MNIST, i.i.d. case.}
	\label{fig:hetro_aware}
	\vspace{-5pt}
\end{figure}

\begin{figure}[tb!]
	\vspace{0pt}
	\centering
	\includegraphics[width =2.65in]{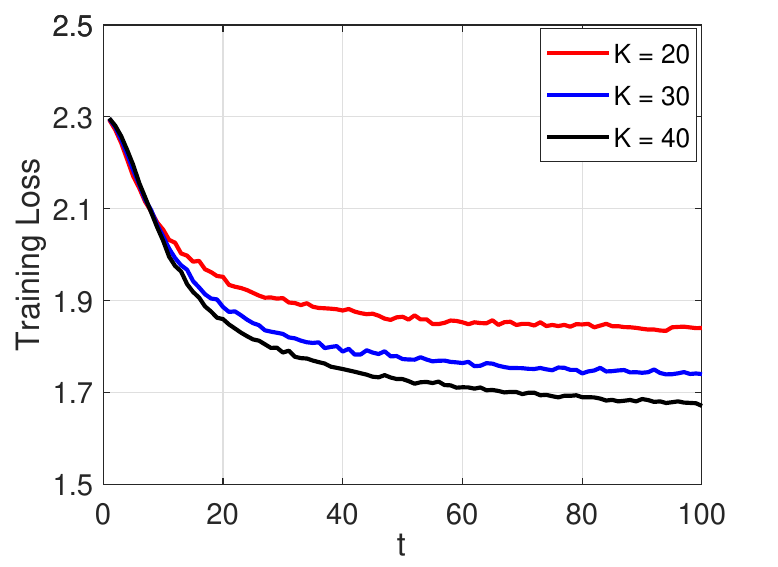} 
	\vspace{-5pt}
	\caption{Algorithm 1, MNIST, non-i.i.d. case.}
	\label{fig:K}
	\vspace{-5pt}
\end{figure}

\begin{figure}[tb!]
	\vspace{0pt}
	\centering
	\includegraphics[width =2.65in]{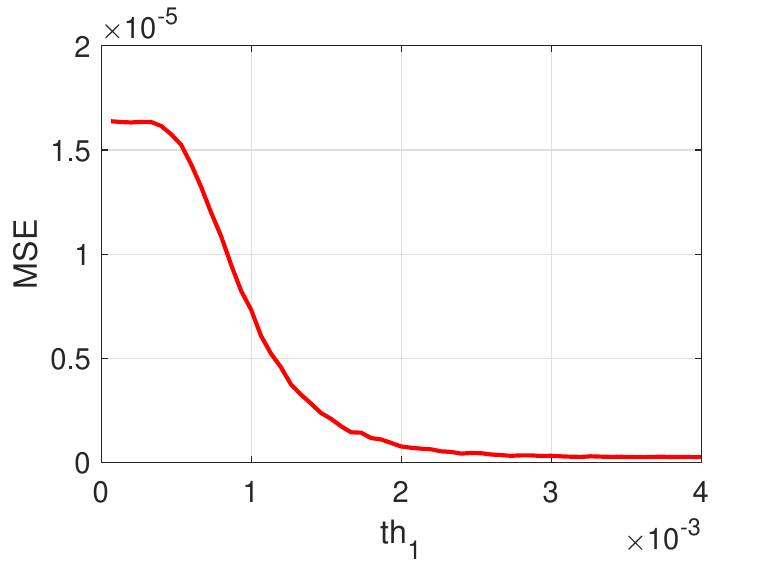} 
	\vspace{-5pt}
	\caption{Algorithm 1, MNIST, non-i.i.d. case.}
	\label{fig:mse}
	\vspace{-5pt}
\end{figure}

\begin{figure}[tb!]
	\vspace{0pt}
	\centering
	\includegraphics[width =2.65in]{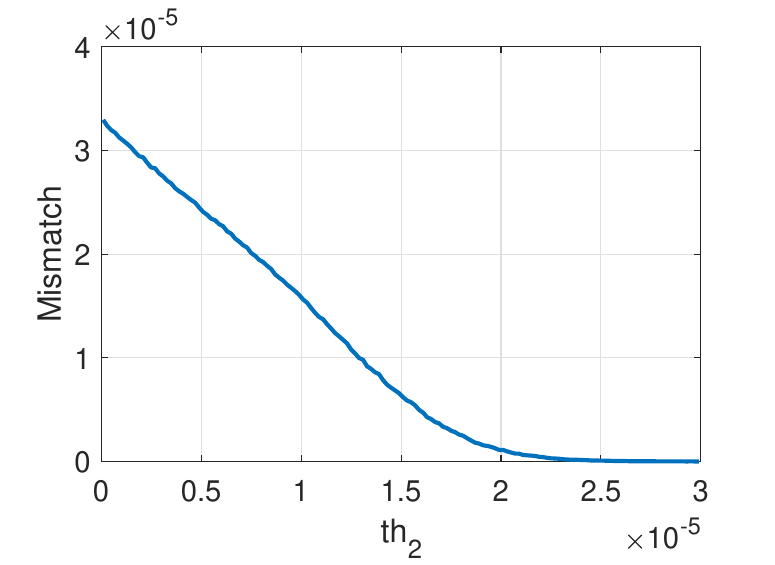} 
	\vspace{-5pt}
	\caption{Algorithm 2, MNIST, non-i.i.d. case.}
	\label{fig:mis}
	\vspace{-5pt}
\end{figure}

\begin{figure}[tb!]
	\vspace{0pt}
	\centering
	\includegraphics[width =2.65in]{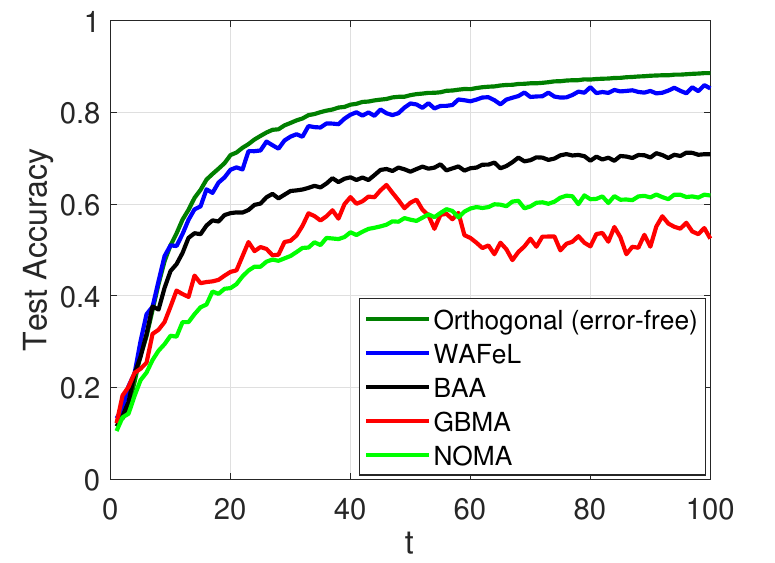} 
	\vspace{-5pt}
	\caption{Algorithm 1, MNIST, non-i.i.d. case.}
	\label{fig:compare}
	\vspace{-5pt}
\end{figure}

\begin{figure}[tb!]
	\vspace{0pt}
	\centering
	\includegraphics[width =2.65in]{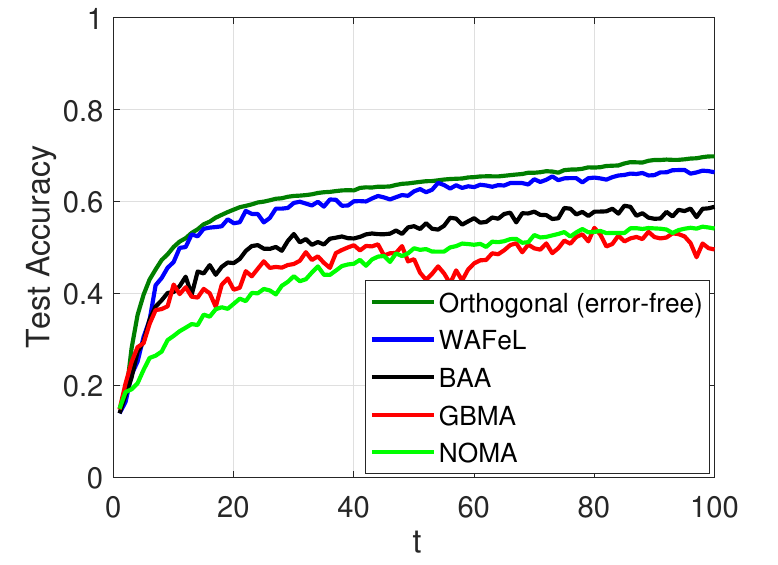} 
	\vspace{-5pt}
	\caption{Algorithm 1, CIFAR-10, non-i.i.d. case.}
	\label{fig:cifar}
	\vspace{-5pt}
\end{figure}


In Fig. \ref{fig:tau}, the training loss is shown for different local iterations $\tau$ in the MNIST and i.i.d. scenario. As observed, increasing $\tau$ or $t$ improves the learning performance. It further exhibits a marked improvement when integrating multiple local iterations in comparison to using just a single local iteration.

In Fig. \ref{fig:hetro_aware}, the test accuracy is shown for both the heterogeneity-aware design of {\fontfamily{lmtt}\selectfont
	WAFeL} and its heterogeneity-unaware design when all devices are set to use an equal batch size based on the capabilities of the least powerful device as the straggler, i.e., with the lowest batch size. 
It is observed that the heterogeneity-aware design significantly outperforms in terms of accuracy, even under the considered limited heterogeneity.

In Fig. \ref{fig:K}, the training loss is shown for different numbers of devices $K$ in the MNIST and non-i.i.d. scenario. The performance improves as $K$ increases because more devices participates in the learning process. However, this improvement comes at the cost of increased MSE, resulting in a tradeoff. Specifically, for higher values of $K$, this performance improvement tends to diminish.

In Figs \ref{fig:mse} and \ref{fig:mis}, the effects of the thresholds, $\text{th}_1$ and $\text{th}_2$, are investigated on the objectives in problems \eqref{practical_obj1} and \eqref{practical_obj2}, the MSE and mismatch, respectively. As the thresholds increase, it is observed that the objectives decrease significantly.

In Figs \ref{fig:compare} and \ref{fig:cifar}, comparisons are made between {\fontfamily{lmtt}\selectfont
	WAFeL} and well-known schemes from the literature, serving as benchmarks, using both MNIST and CIFAR-10 datasets in the non-i.i.d. scenario. Among the benchmarks using analog over-the-air computation similar to {\fontfamily{lmtt}\selectfont
	WAFeL} are the {\fontfamily{lmtt}\selectfont
	BAA} scheme, which uses truncated power allocation with accurate knowledge of CSIT \cite{huang_analog}, and the {\fontfamily{lmtt}\selectfont
	GBMA} scheme, which compensates for phase only at the transmitters \cite{cohen1}. For both {\fontfamily{lmtt}\selectfont
	BAA} and {\fontfamily{lmtt}\selectfont
	GBMA}, perfect fine synchronization is considered. Additionally, the ideal error-free performance is considered using FL with digital orthogonal transmissions, requiring at least $K = 30$ times more resource blocks. The FL using NOMA \cite{noma}, referred to as {\fontfamily{lmtt}\selectfont
	NOMA}, is also considered as another benchmark via digital communications, under the same single resource block as in {\fontfamily{lmtt}\selectfont
	WAFeL}, {\fontfamily{lmtt}\selectfont
	BAA}, and {\fontfamily{lmtt}\selectfont
	GBMA}. For a consistent comparison, {\fontfamily{lmtt}\selectfont
	FedAvg} with the same $\tau = 3$ is implemented across all the schemes. While the {\fontfamily{lmtt}\selectfont
	BAA} scheme possesses a distinct complexity level owing to its need for perfect CSIT and channel compensation requirements, the blind {\fontfamily{lmtt}\selectfont
	GBMA} method lacks any optimization algorithm or processing, setting it apart from {\fontfamily{lmtt}\selectfont
	WAFeL} which incorporates an algorithm of complexity $\mathcal{O}(K^3)$. 

It is observed that {\fontfamily{lmtt}\selectfont
	WAFeL} significantly outperforms both the {\fontfamily{lmtt}\selectfont
	BAA} and {\fontfamily{lmtt}\selectfont
	GBMA} schemes across both the MNIST and CIFAR-10 datasets. For example, at $t = 100$, the improvement of {\fontfamily{lmtt}\selectfont
	WAFeL} is around $15 \%$ and $30\%$ compared to {\fontfamily{lmtt}\selectfont
	BAA} and {\fontfamily{lmtt}\selectfont
	GBMA}, respectively, when evaluated with the MNIST dataset. Moreover, it closely approximates the performance achieved by the orthogonal case. Surprisingly, {\fontfamily{lmtt}\selectfont
	WAFeL}, which incorporates partial phase compensation, outperforms schemes like {\fontfamily{lmtt}\selectfont
	BAA} that have perfect gain and phase compensation. This performance improvement is attributed to the utilization of optimized adaptive weights in the aggregation process, which was not previously identified, coupled with the unique receiver structure. It is noteworthy that in contrast to {\fontfamily{lmtt}\selectfont
	BAA} where each device needs to adhere to both average and maximum power constraints for power allocation and there is an inherent device selection, the weight allocation in {\fontfamily{lmtt}\selectfont
	WAFeL} is not subjected to any limitations and all the devices contribute to the learning. 

Also, {\fontfamily{lmtt}\selectfont
	WAFeL}, which is based on analog communications, significantly outperforms {\fontfamily{lmtt}\selectfont
	NOMA}, which is based on digital communications. This is mainly because in {\fontfamily{lmtt}\selectfont
	NOMA}, not only is interference considered as additional noise when decoding individual model parameters of the devices, but also model parameters must be quantized into finite bits limited by the available resource block before transmission, further adding quantization errors. On the other hand, in {\fontfamily{lmtt}\selectfont
	WAFeL}, the aggregation is estimated directly, and aggregation weights are optimally designed to maximize convergence performance.


\section{Conclusions}
In this paper, we introduced a unique over-the-air federated learning scheme incorporating a novel weighted aggregation approach. In the scheme, the adaptive choice of aggregation weights helps counteract the impacts of wireless channels on performance, eliminating the need for channel compensation at the transmitting end. By considering device heterogeneity via diverse batch sizes, we analyzed the convergence rate of the learning process for the scheme as a function of the aggregation weights, which also includes both communication and learning factors. To select the aggregation weights, we proposed aggregation cost metrics based on the analysis. We presented efficient algorithms to optimize the metrics, and also derived an achievable bound on the convergence rate. Despite channel conditions and device heterogeneity, experimental results demonstrated the high learning accuracy of the proposed scheme, surpassing existing solutions even the one with channel compensation at transmitters. Moreover, the proposed scheme can closely attain the level of performance that would be expected in the absence of any errors. Consequently, the proposed scheme, which emphasizes aggregation weight optimization, offers a promising new learning framework that can be further explored in various setups for its potential applications.

\appendices
\section{Proof of Theorem 1}
The update of learning model at round $t+1$ is as
\begin{align} 
&{\mathbf{w}}_{\text{G},t+1}^\top = \frac{1}{\sqrt{P}}\mathbf{b}_t^\top \mathbf{Y}_t+\sum_{k=1}^{K}\alpha_{k,t}\mu_{k,t}\mathbf{1}^\top = \sum_{k=1}^{K}\alpha_{k,t}\mathbf{w}_{k,t}^\top+\nonumber\\
&{\boldsymbol\epsilon(\boldsymbol\alpha_t)}= {\mathbf{w}}_{\text{G},t}^\top -{\eta}\sum_{k=1}^{K}\alpha_{k,t} \sum_{i=0}^{\tau-1}\nabla F_k(\mathbf{w}_{k,t,i},\boldsymbol\xi_{k}^i)^\top+{\boldsymbol\epsilon(\boldsymbol\alpha_t)},
\end{align}
where $\boldsymbol\epsilon(\boldsymbol{\alpha}_t)$ models the estimation error in recovering the aggregation with the weight vector $\boldsymbol{\alpha}_t$. Subsequently, based on the $L$-Lipschitz continuity attribute stated in Assumption 1, we obtain
\begin{align}
\label{lipsch}
&F({\mathbf{w}}_{\text{G},t+1}) - F({\mathbf{w}}_{\text{G},t}) \leq \nabla F( {\mathbf{w}}_{\text{G},t})^\top \left({\mathbf{w}}_{\text{G},t+1} -  {\mathbf{w}}_{\text{G},t}\right)+\nonumber\\
&\frac{L}{2} \| {\mathbf{w}}_{\text{G},t+1} - {\mathbf{w}}_{\text{G},t}\|^2 =\nabla F( {\mathbf{w}}_{\text{G},t})^\top \biggl(-{\eta}\sum_{k=1}^{K}\alpha_{k,t}\nonumber\\& \sum_{i=0}^{\tau-1}\nabla F_k(\mathbf{w}_{k,t,i},\boldsymbol\xi_{k}^i)+{\boldsymbol\epsilon(\boldsymbol\alpha_t)}\biggr)+\frac{L}{2}\times \nonumber\\
&\left\Vert -{\eta}\sum_{k=1}^{K}\alpha_{k,t} \sum_{i=0}^{\tau-1}\nabla F_k(\mathbf{w}_{k,t,i},\boldsymbol\xi_{k}^i)+{\boldsymbol\epsilon(\boldsymbol\alpha_t)}\right\Vert^2.
\end{align}
By applying expectation to both sides of \eqref{lipsch}, we proceed as
\begin{align}
\label{expected}
&\mathbb{E}\left\{F( {\mathbf{w}}_{\text{G},t+1}) - F( {\mathbf{w}}_{\text{G},t}) \right\} \leq -{\eta}\sum_{k=1}^{K}\alpha_{k,t} \sum_{i=0}^{\tau-1} \nonumber\\
&\mathbb{E}\left\{\nabla F( {\mathbf{w}}_{\text{G},t})^\top \nabla F_k(\mathbf{w}_{k,t,i},\boldsymbol\xi_{k}^i)\right\}+\frac{L\eta^2}{2} \mathbb{E}\Biggl\{ \Biggl\Vert \sum_{k=1}^{K}{\alpha_{k,t}} \sum_{i=0}^{\tau-1}\nonumber\\
&\nabla F_k(\mathbf{w}_{k,t,i},\boldsymbol\xi_{k}^i)\Biggr\Vert^2\Biggr\} +\frac{L}{2}\mathbb{E}\left\{\left\Vert {\boldsymbol\epsilon(\boldsymbol\alpha_t)}\right\Vert^2\right\}= -{\eta}\sum_{k=1}^{K}\alpha_{k,t}\nonumber\\
& \sum_{i=0}^{\tau-1} \mathbb{E}\left\{\nabla F( {\mathbf{w}}_{\text{G},t})^\top \nabla F(\mathbf{w}_{k,t,i})\right\}+\frac{L\eta^2}{2} \mathbb{E}\Biggl\{ \Biggl\Vert \sum_{k=1}^{K}{\alpha_{k,t}}\nonumber\\
& \sum_{i=0}^{\tau-1}\nabla F_k(\mathbf{w}_{k,t,i},\boldsymbol\xi_{k}^i)\Biggr\Vert^2\Biggr\} +\frac{L}{2} {\text{MSE}(\boldsymbol\alpha_t)},
\end{align}
where the fact $\mathbb{E}\left\{\boldsymbol\epsilon(\boldsymbol{\alpha}_t)\right\} = \mathbf{0}$ is considered, due to the transmit normalization. Next, we bound the first term of the right-hand side (RHS) in \eqref{expected}. We can write its inner-sum
term, using the equation $\|\mathbf{t}_1-\mathbf{t}_2\|^2 = \|\mathbf{t}_1\|^2+\|\mathbf{t}_2\|^2- 2\mathbf{t}_1^\top \mathbf{t}_2$, as 
\begin{align}
\label{diffexp}
&\mathbb{E}\left\{\nabla F( {\mathbf{w}}_{\text{G},t})^\top \nabla F(\mathbf{w}_{k,t,i})\right\} = \frac{1}{2} \mathbb{E}\left\{\|\nabla F( {\mathbf{w}}_{\text{G},t})\|^2\right\} +\nonumber\\
& \frac{1}{2} \mathbb{E}\left\{\|\nabla F(\mathbf{w}_{k,t,i})\|^2\right\} - \frac{1}{2} \mathbb{E}\left\{\|\nabla F( {\mathbf{w}}_{\text{G},t})- \nabla F(\mathbf{w}_{k,t,i})\|^2\right\}.
\end{align}
From Assumption 1, the last term in \eqref{diffexp} is bounded as
\begin{align}
\label{diffnorm}
&\mathbb{E}\left\{\|\nabla F( {\mathbf{w}}_{\text{G},t})- \nabla F(\mathbf{w}_{k,t,i})\|^2\right\} \leq L^2 \mathbb{E}\left\{\|\mathbf{w}_{\text{G},t}-\mathbf{w}_{k,t,i}\|^2\right\}\nonumber\\
&= L^2 \mathbb{E}\Biggl\{\Biggl\Vert-\eta \sum_{j=0}^{i-1} \nabla F_k(\mathbf{w}_{k,t,j},\boldsymbol\xi_{k}^j)\Biggr \Vert^2\Biggr\}\nonumber\\
&=L^2\eta^2 \mathbb{E}\Biggl\{\Biggl\Vert\sum_{j=0}^{i-1}\nabla F_k(\mathbf{w}_{k,t,j},\boldsymbol\xi_{k}^j) \Biggr\Vert^2\Biggr\},
\end{align}
where, by employing the equation $\mathbb{E}\left\{\|\mathbf{t}\|^2\right\} = \|\mathbb{E}\left\{\mathbf{t}\right\}\|^2 + \mathbb{E}\left\{\|\mathbf{t}-\mathbb{E}\left\{\mathbf{t}\right\}\|^2\right\}$, we derive the following
\begin{align}
\label{normexp}
&\mathbb{E}\Biggl\{\Biggl\Vert\sum_{j=0}^{i-1}\nabla F_k(\mathbf{w}_{k,t,j},\boldsymbol\xi_{k}^j) \Biggr\Vert^2\Biggr\} =\mathbb{E}\Biggl\{\Biggl\Vert\sum_{j=0}^{i-1}\nabla F(\mathbf{w}_{k,t,j}) \Biggr\Vert^2\Biggr\}\nonumber\\
&+\mathbb{E}\Biggl\{\Biggl\Vert \sum_{j=0}^{i-1} \nabla F_k(\mathbf{w}_{k,t,j},\boldsymbol\xi_{k}^j)- \nabla F(\mathbf{w}_{k,t,j})\Biggr\Vert^2\Biggr\},
\end{align}
where the first term of RHS can be upper-bounded as
\begin{align}
\label{RHS1}
\mathbb{E}\Biggl\{\Biggl\Vert\sum_{j=0}^{i-1}\nabla F(\mathbf{w}_{k,t,j}) \Biggr\Vert^2\Biggr\} &\stackrel{(a)}{\leq}  i \sum_{j=0}^{i-1}\mathbb{E}\left\{\left\Vert\nabla F(\mathbf{w}_{k,t,j}) \right\Vert^2\right\},
\end{align}
where $(a)$ comes from the inequality
of arithmetic and geometric means, i.e., $\left(\sum_{i=1}^{I}a_i\right)^2\leq I \sum_{i=1}^{I}a_i^2$. The second term of RHS in \eqref{normexp} can be upper-bounded as
\begin{align}
\label{RHS2}
&\mathbb{E}\Biggl\{\Biggl\Vert \sum_{j=0}^{i-1} \nabla F_k(\mathbf{w}_{k,t,j},\boldsymbol\xi_{k}^j)- \nabla F(\mathbf{w}_{k,t,j})\Biggr\Vert^2\Biggr\}\stackrel{(b)}{=}\nonumber\\
&  \sum_{j=0}^{i-1}\mathbb{E}\left\{\left\Vert \nabla F_k(\mathbf{w}_{k,t,j},\boldsymbol\xi_{k}^j)- \nabla F(\mathbf{w}_{k,t,j})\right\Vert^2\right\}
\stackrel{(c)}{\leq} i\frac{\sigma_\text{g}^2}{B_k},
\end{align}
where $(b)$ is due to the independence conditioned on batches $\boldsymbol\xi_{k}^j$ for any two distinct values of $k$ or $j$. 
Also, $(c)$ comes from Assumption 2. Replacing \eqref{RHS1} and \eqref{RHS2} in \eqref{normexp} and then replacing the result in \eqref{diffnorm}, we have
\begin{align}
\label{product}
&\mathbb{E}\left\{\nabla F( {\mathbf{w}}_{\text{G},t})^\top \nabla F(\mathbf{w}_{k,t,i})\right\} \geq \frac{1}{2} \mathbb{E}\left\{\|\nabla F( {\mathbf{w}}_{\text{G},t})\|^2\right\} + \nonumber\\
&\frac{1}{2} \mathbb{E}\left\{\|\nabla F(\mathbf{w}_{k,t,i})\|^2\right\} - \frac{L^2 \eta^2}{2}i \sum_{j=0}^{i-1}\mathbb{E}\left\{\left\Vert\nabla F(\mathbf{w}_{k,t,j}) \right\Vert^2\right\}\nonumber\\
&-\frac{L^2\eta^2}{2}i\frac{\sigma_\text{g}^2}{B_k}.
\end{align}
Therefore, we obtain the following bound
\begin{align}
\label{RHS1of}
&-{\eta}\sum_{k=1}^{K}\alpha_{k,t} \sum_{i=0}^{\tau-1} \mathbb{E}\left\{\nabla F( {\mathbf{w}}_{\text{G},t})^\top \nabla F(\mathbf{w}_{k,t,i})\right\} \leq  -\frac{\eta\tau}{2}\times \nonumber\\& \mathbb{E}\left\{\|\nabla F( {\mathbf{w}}_{\text{G},t})\|^2\right\}-\frac{\eta}{2}\sum_{k=1}^{K}\alpha_{k,t} \sum_{i=0}^{\tau-1} \mathbb{E}\left\{\|\nabla F(\mathbf{w}_{k,t,i})\|^2\right\}+\nonumber\\
&\frac{L^2\eta^3}{2}\sum_{k=1}^{K}\alpha_{k,t} \sum_{i=0}^{\tau-1} i \sum_{j=0}^{i-1}\mathbb{E}\left\{\left\Vert\nabla F(\mathbf{w}_{k,t,j}) \right\Vert^2\right\}+\nonumber\\&\frac{L^2\eta^3}{2}\frac{\tau (\tau-1)}{2}{\sigma_\text{g}^2} \sum_{k=1}^{K}\frac{\alpha_{k,t}}{B_k}.
\end{align}
Next, we bound the second term of the RHS in \eqref{expected} as
\begin{align}
\label{weighted_norm}
&\mathbb{E}\Biggl\{ \Biggl\Vert \sum_{k=1}^{K}{\alpha_{k,t}} \sum_{i=0}^{\tau-1}\nabla F_k(\mathbf{w}_{k,t,i},\boldsymbol\xi_{k}^i)\Biggr\Vert^2\Biggr\} = \nonumber\\
&\mathbb{E}\Biggl\{ \Biggl\Vert \sum_{k=1}^{K}{\alpha_{k,t}} \sum_{i=0}^{\tau-1}\nabla F(\mathbf{w}_{k,t,i})\Biggr\Vert^2\Biggr\}+\nonumber\\
&\mathbb{E}\Biggl\{ \Biggl\Vert \sum_{k=1}^{K}{\alpha_{k,t}} \sum_{i=0}^{\tau-1}\left(\nabla F_k(\mathbf{w}_{k,t,i},\boldsymbol\xi_{k}^i)-\nabla F(\mathbf{w}_{k,t,i})\right)\Biggr\Vert^2\Biggr\},
\end{align}
where
\begin{align}
\label{RRHS1}
&\mathbb{E}\Biggl\{ \Biggl\Vert \sum_{k=1}^{K}{\alpha_{k,t}} \sum_{i=0}^{\tau-1}\nabla F(\mathbf{w}_{k,t,i})\Biggr\Vert^2\Biggr\} \nonumber\\
&\stackrel{(d)}{\leq} \sum_{k=1}^{K}{\alpha_{k,t}} \mathbb{E}\Biggl\{ \Biggl\Vert\sum_{i=0}^{\tau-1}\nabla F(\mathbf{w}_{k,t,i})\Biggr\Vert^2\Biggr\} \nonumber\\
&\stackrel{(e)}{\leq} \tau \sum_{k=1}^{K}{\alpha_{k,t}} \sum_{i=0}^{\tau-1}\mathbb{E}\left\{ \left\Vert\nabla F(\mathbf{w}_{k,t,i})\right\Vert^2\right\}, 
\end{align}
where $(d)$ follows from the convexity of $\|.\|^2$ and $(e)$ is from the inequality of arithmetic
and geometric means. Also, the second term of RHS in \eqref{weighted_norm} can be upper-bounded as
\begin{align}
\label{RRHS2}
&\mathbb{E}\Biggl\{ \Biggl\Vert \sum_{k=1}^{K}{\alpha_{k,t}} \sum_{i=0}^{\tau-1}\left(\nabla F_k(\mathbf{w}_{k,t,i},\boldsymbol\xi_{k}^i)-\nabla F(\mathbf{w}_{k,t,i})\right)\Biggr\Vert^2\Biggr\} =\nonumber\\
& \sum_{k=1}^{K}{\alpha_{k,t}^2} \sum_{i=0}^{\tau-1}\mathbb{E}\left\{ \left\Vert\nabla F_k(\mathbf{w}_{k,t,i},\boldsymbol\xi_{k}^i)-\nabla F(\mathbf{w}_{k,t,i})\right\Vert^2\right\} \nonumber\\&\leq {\sigma_\text{g}^2}\tau \sum_{k=1}^{K}\frac{\alpha_{k,t}^2}{B_k},
\end{align}
which comes from the conditional independence and Assumption 2. Now, replacing \eqref{RRHS1} and \eqref{RRHS2} in \eqref{weighted_norm} and replacing the result with \eqref{RHS1of} in \eqref{expected}, we obtain
\begin{align}
&\mathbb{E}\left\{F( {\mathbf{w}}_{\text{G},t+1}) - F( {\mathbf{w}}_{\text{G},t}) \right\} \leq  -\frac{\eta\tau}{2} \mathbb{E}\left\{\|\nabla F( {\mathbf{w}}_{\text{G},t})\|^2\right\}\nonumber\\&-\frac{\eta}{2}\sum_{k=1}^{K}\alpha_{k,t} \sum_{i=0}^{\tau-1} \mathbb{E}\left\{\|\nabla F(\mathbf{w}_{k,t,i})\|^2\right\}+\frac{L^2\eta^3}{2}\sum_{k=1}^{K}\alpha_{k,t} \nonumber\\
&\sum_{i=0}^{\tau-1} i \sum_{j=0}^{i-1}\mathbb{E}\left\{\left\Vert\nabla F(\mathbf{w}_{k,t,j}) \right\Vert^2\right\}+\frac{L^2\eta^3}{2}\frac{\tau (\tau-1)}{2}{\sigma_\text{g}^2} \sum_{k=1}^{K}\frac{\alpha_{k,t}}{B_k}\nonumber\\
&+\frac{L\eta^2}{2}\tau \sum_{k=1}^{K}{\alpha_{k,t}} \sum_{i=0}^{\tau-1}\mathbb{E}\left\{ \left\Vert\nabla F(\mathbf{w}_{k,t,i})\right\Vert^2\right\}+\nonumber\\
&\frac{L \eta^2}{2}{\sigma_\text{g}^2}\tau \sum_{k=1}^{K}\frac{{\alpha_{k,t}^2}}{B_k}  +\frac{L}{2} {\text{MSE}(\boldsymbol\alpha_t)}, 
\end{align}
which, when utilizing the bound $\sum_{i=0}^{\tau-1} i \sum_{j=0}^{i-1}\\ \mathbb{E}\left\{\left\Vert\nabla F(\mathbf{w}_{k,t,j}) \right\Vert^2\right\} \leq \sum_{i=0}^{\tau-1} i \times \sum_{i=0}^{\tau-1} \mathbb{E}\left\{\left\Vert\nabla F(\mathbf{w}_{k,t,i}) \right\Vert^2\right\} \\= \frac{\tau(\tau-1)}{2}\sum_{i=0}^{\tau-1} \mathbb{E}\left\{\left\Vert\nabla F(\mathbf{w}_{k,t,i}) \right\Vert^2\right\}$, can be upper-bounded as
\begin{align}
&\mathbb{E}\left\{F( {\mathbf{w}}_{\text{G},t+1}) - F( {\mathbf{w}}_{\text{G},t}) \right\} \leq  -\frac{\eta\tau}{2} \mathbb{E}\left\{\|\nabla F( {\mathbf{w}}_{\text{G},t})\|^2\right\}-\nonumber\\
&\frac{\eta}{2}\sum_{k=1}^{K}\alpha_{k,t} \sum_{i=0}^{\tau-1} \mathbb{E}\left\{\|\nabla F(\mathbf{w}_{k,t,i})\|^2\right\}+\frac{L^2\eta^3}{2}\frac{\tau(\tau-1)}{2}\nonumber\\
&\sum_{k=1}^{K}\alpha_{k,t} \sum_{i=0}^{\tau-1} \mathbb{E}\left\{\left\Vert\nabla F(\mathbf{w}_{k,t,i}) \right\Vert^2\right\}+\frac{L^2\eta^3}{2}\frac{\tau (\tau-1)}{2}{\sigma_\text{g}^2} \nonumber\\
&\sum_{k=1}^{K}\frac{\alpha_{k,t}}{B_k}+\frac{L\eta^2}{2}\tau \sum_{k=1}^{K}{\alpha_{k,t}} \sum_{i=0}^{\tau-1}\mathbb{E}\left\{ \left\Vert\nabla F(\mathbf{w}_{k,t,i})\right\Vert^2\right\}+\frac{L \eta^2}{2}\nonumber\\
& {\sigma_\text{g}^2}\tau\sum_{k=1}^{K}\frac{{\alpha_{k,t}^2}}{B_k}  +\frac{L}{2} {\text{MSE}(\boldsymbol\alpha_t)} =  -\frac{\eta}{2}\biggl(1-\frac{L^2\eta^2}{2}\tau(\tau-1)-\nonumber\\
&L\eta \tau\biggr)\sum_{k=1}^{K}{\alpha_{k,t}} \sum_{i=0}^{\tau-1} \mathbb{E}\left\{\|\nabla F(\mathbf{w}_{k,t,i})\|^2\right\}+\frac{L^2\eta^3}{2}\frac{\tau (\tau-1)}{2}{\sigma_\text{g}^2} \nonumber\\
&\sum_{k=1}^{K}\frac{\alpha_{k,t}}{B_k}+\frac{L \eta^2}{2}{\sigma_\text{g}^2}\tau \sum_{k=1}^{K}\frac{{\alpha_{k,t}^2}}{B_k} +\frac{L}{2} {\text{MSE}(\boldsymbol\alpha_t)}\nonumber\\
&-\frac{\eta\tau}{2} \mathbb{E}\left\{\|\nabla F( {\mathbf{w}}_{\text{G},t})\|^2\right\}.
\end{align}
Thus, under the condition
\begin{align}
1-\frac{L^2\eta^2}{2}\tau(\tau-1)-L\eta \tau \geq 0,
\end{align}
we have
\begin{align}
&\mathbb{E}\left\{F( {\mathbf{w}}_{\text{G},t+1}) - F( {\mathbf{w}}_{\text{G},t}) \right\} \leq  -\frac{\eta\tau}{2} \mathbb{E}\left\{\|\nabla F( {\mathbf{w}}_{\text{G},t})\|^2\right\}+\nonumber\\
&\frac{L^2\eta^3}{2}\frac{\tau (\tau-1)}{2}{\sigma_\text{g}^2} \sum_{k=1}^{K}\frac{\alpha_{k,t}}{B_k}+\frac{L \eta^2}{2}{\sigma_\text{g}^2}\tau \sum_{k=1}^{K}\frac{\alpha_{k,t}^2}{B_k} +\frac{L}{2} {\text{MSE}(\boldsymbol\alpha_t)}.
\end{align}
After performing a telescoping sum over the global rounds $t \in \left\{0,\ldots,T-1\right\}$, we obtain
\begin{align}
\label{finalstep}
&\mathbb{E}\left\{F({\mathbf{w}}_{\text{G},T})\right\}-F({\mathbf{w}}_{\text{G},0}) \leq -\frac{\eta\tau}{2} \sum_{t=0}^{T-1}\mathbb{E}\left\{\|\nabla F( {\mathbf{w}}_{\text{G},t})\|^2\right\}+\nonumber\\&\sum_{t=0}^{T-1}\biggl(\frac{L^2\eta^3}{2}\frac{\tau (\tau-1)}{2}{\sigma_\text{g}^2} \sum_{k=1}^{K}\frac{\alpha_{k,t}}{B_k}+\frac{L \eta^2}{2}{\sigma_\text{g}^2}\tau \sum_{k=1}^{K}\frac{\alpha_{k,t}^2}{B_k} \nonumber\\&+\frac{L}{2} {\text{MSE}(\boldsymbol\alpha_t)}\biggr),
\end{align}
and then using the fact $\mathbb{E}\left\{F(\mathbf{w}_{\text{G},T})\right\}\geq F^{*}$, we reach the conclusion of the proof.

\end{document}